\documentclass[aps,prd,nofootinbib]{revtex4}
\usepackage{graphicx}
\usepackage{amsfonts}
\usepackage{amsmath}
\usepackage{amssymb}

\begin{document}

\title{Galactic microlensing by backreacted massless wormholes}
\author{G.F. Akhtaryanova}
\email{akht\_gul@mail.ru}
\affiliation{Zel'dovich International Center for Astrophysics, Bashkir State Pedagogical University, 3A, October Revolution Street, Ufa 450008, RB, Russia}
\author{R.Kh. Karimov}
\email{karimov_ramis_92@mail.ru}
\affiliation{Zel'dovich International Center for Astrophysics, Bashkir State Pedagogical University, 3A, October Revolution Street, Ufa 450008, RB, Russia}
\author{R.N. Izmailov}
\email{izmailov.ramil@gmail.com}
\affiliation{Zel'dovich International Center for Astrophysics, Bashkir State Pedagogical University, 3A, October Revolution Street, Ufa 450008, RB, Russia}
\author{K.K. Nandi}
\email{kamalnandi1952@rediffmail.com}
\affiliation{Zel'dovich International Center for Astrophysics, Bashkir State Pedagogical University, 3A, October Revolution Street, Ufa 450008, RB, Russia}
\affiliation{High Energy Cosmic Ray Research Center, University of North Bengal, Siliguri 734013, WB, Bharat}

\date{13 May 2024}

\begin{abstract}
We study here a novel application of Kim \& Lee charged wormholes assuming them to be dark halo objects playing the role of lenses in the Galactic microlensing with source stars belonging to the Galactic Bulge and the Large Magellanic Cloud. First, we observe that both the backreacted scalar ($\alpha $) and electrically ($Q$) charged wormholes have the same zero ADM mass as has the background Ellis-Bronnikov wormhole having a special equation of state parameter $\gamma = - 1$. In particular, we argue that, for $\alpha \neq 0$, the solution formally resembles, but can at best be sourcewise different from, that of the background wormhole. The charge ($Q\neq 0$) thus provides an extra degree of freedom that introduces a non-trivial redshift function $\Phi$ to the background, alters its throat radius to $r_{\text{th}}$, yet keeps the wormhole massless. Second, we focus on this electrically charged case and calculate the light deflection angle up to 4$^{\text{th}}$ PPN order, analyze the effect of $Q$ on the lensing observables such as the image positions, magnification, centroid and time delay of images of the source stars. Third, we analyze the probabilistic features such as optical depth and event rate estimated on the basis of the hypothesis that the wormhole lens could be bound or unbound to our Galaxy. Finally, we report an intriguing qualitative prediction that, compared to the Schwarzschild black hole, the Paczy\'{n}ski light curves of the electrically charged wormhole are much dimmer that also show characteristic gutters at the times the source enters and exits the Einstein ring. \textit{The gutters gradually come together as $Q$ approaches the extreme limit $\frac{r_{\text{th}}}{\sqrt{2}}$, at which the Einstein radius $R_{E}$ vanishes so that the source crosses it instantly.} It is speculated that re-analyzing past data on Galactic microlensing may betray the presence of charged wormholes.
\end{abstract}

%\pacs{}
\maketitle

%%%%%%%%%%%%%%%%%%  DATE  %%%%%%%%%%%%%%%%%%%

%%%%%%%%%%%%%%%%%%%%%%%%%%%%%%%%%%%%%%%%%
\section{Introduction}\label{sec1}
%%%%%%%%%%%%%%%%%%%%%%%%%%%%%%%%%%%%%%%%%
Wormholes are handles in spacetime with non-trivial topology that connect two distant regions of space or even two universes\footnote{%
Wheeler and Misner first coined the term "wormhole" in \cite{Misner:1957}. Their geometry considers lines of force forming what topologists would call "a handle" of the multiply-connected space, and what physicists would call more picturesquely "a wormhole". We are concerned in this paper only with Lorentzian wormholes with signature ($-,+,+,+$).}. Wormhole spacetimes are valid solutions of Einstein's general relativity that have not yet been ruled out by experiments. An early (1918) predecessor of wormholes is the Flamm paraboloid \cite{Rindler:1979}. Then, after the lapse of several years, Einstein and Rosen \cite{Einstein:1935} proposed in 1935 a particle model for massless electric charge in the form of a bridge (wormhole throat). Still about half a century later, the concept of wormholes has been systematically formalized by Morris and Thorne in their pioneering work of 1988 \cite{Morris:1988a}. Though wormholes were presented in \cite{Morris:1988a} as a curious means of interstellar travel by way of a pedagogical tool for teaching general relativity, it is very much a part of frontier physics today since the scientific community was quick to realize that the formalism could be applied to much wider regimes encompassing phenomenologies from quantum to classical general relativity \cite{Morris:1988b, Visser:1995}. One such important regime concerns gravitational lensing though it is practically impossible to exhaustively cite the voluminous works on this topic that have collected over past several decades. We anyway randomly cite a handful \cite{Schneider:1992, Bambi:2021, Bozza:2001, Bozza:2002, Virbhadra:2008, Frittelli:2000, Virbhadra:2000, Virbhadra:2002, Keeton:2005}. For an early classic review on lensing by compact objects, see \cite{Schneider:1992} and for a recent succint review on the possibility of astrophysical wormholes, see \cite{Bambi:2021}. For example, Kardashev, Novikov and Shatskiy \cite{Kardashev:2006} suggest that some active galactic nuclei and other compact astrophysical objects may be interpreted as wormholes. Keeton and Petters \cite{Keeton:2005} developed PPN expansion method for finding weak field lensing observables, which we shall use here, and Bozza \cite{Bozza:2002} developed those for strong field lensing, when the light rays pass arbitrarily close to the photon sphere, if it exists.

While gravitational lensing is quite a powerful diagnostic that can be used to distinguish between compact objects such as black holes and/or naked singularities \cite{Virbhadra:2000, Virbhadra:2002, Izmailov:2020}, the diagnostic can also potentially distinguish between wormholes and compact objects \cite{Kardashev:2006, Tsukamoto:2012, Karimov:2020, Lukmanova:2018, Izmailov:2019, Bronnikov:2019, Asada:2017, Shaikh:2019, Yoo:2013, Ishkaeva:2023, Nandi:2018, Kasuya:2021, Cheng:2021, Bugaev:2021, Godani:2023, Perlick:2023, Harko:2009, Cramer:1995, Safonova:2001, Eiroa:2001, Chetouani:1984, Toki:2011, Nakajima:2012}. Apart from lensing signatures \cite{Rahaman:2023, Nandi:2006, Abe:2010, Bhattacharya:2010}, there could be other distinctive properties of wormholes. For instance, wormholes resulting from the merger of two Schwarzschild black holes can emit gravitational waves \cite{Cardoso:2016}. Such ring-down emissions can occur also from the perturbations of massive Ellis-Bronnikov wormholes \cite{Nandi:2017}. Wormholes can leave distinguishing imprints on Paczy\'{n}ski light curves \cite{Abe:2010, Paczynski:1986}, on the accretion disk profiles \cite{Karimov:2019, Yusupova:2021} - to name a few. Wormholes appear also as exact solutions of other competing theories, for example, in the Machian Brans-Dicke theory \cite{Agnese:1995, Anchordoqui:1997, Nandi:1997, Izmailov:2020-2} as well as in the Einstein conformal frame in which the Brans-Dicke theory converts to Einstein minimally coupled scalar field theory \cite{Nandi:1998}.

Kim and Lee \cite{Kim:2001} charged wormholes are of particular interest since the backreaction of the additional matter can modify the background spacetime yielding new redshift and shape functions containing extra degrees of freedom in the form of scalar or electric charges (see also \cite{Kim:1996, Kim:1998}). These charged wormholes have been analyzed in the literature from various perspectives. Most recently, the electric charged case ($Q\neq 0$) has been studied by Gao et al \cite{Gao:2023}, who calculated the angle of deflection of light up to second order by using the Gauss-Bonnet method and also the magnification of images. The shadow of the charged wormhole has been studied by Neto et al \cite{Neto:2023}. Modified charged wormhole was proposed in \cite{Jusufi:2019}, which was shown to satisfy quantum Ford-Roman inequality. Non-commutative geometry-inspired charged wormholes with low tidal forces have been studied by Kuhfittig and Gladney \cite{Kuhfittig:2017}.

In the Galactic microlensing scenario, the backreacted wormholes could play the role of re-defined lenses that take into account the effect of additional matter distribution on the background spacetime which, in the present paper, is the electrically neutral massless background Ellis-Bronnikov wormhole (let's call it clean lens). An important point to note is that, even if the ADM mass is zero, the spacetime can still exhibit gravitational effects due to non-zero spacetime curvature. Microlensing properties of this background wormhole have been studied in a seminal paper by Abe \cite{Abe:2010}. Thus the effect of additional charges on the clean lens is expected to be manifest in the lensing signatures of the backreacted geometry, particularly in the Paczy\'{n}ski light curves. This exciting possibility motivated us to explore the precise effects of the electric charge on the microlensing observables in an astrophysically natural setting.

The purpose of this paper is to consider, with Kim and Lee, the special case of background matter equation of state parameter $\gamma =-1$ and comprehensively study Galactic microlensing by backreacted charged wormholes, when the source stars are in the Galactic Bulge or in the Large Magellanic Cloud. We shall analyze, partially following Abe \cite{Abe:2010}, the lensing observables such as the light deflection angle, image positions, magnification, centroid and time delay of images including the probabilistic features such as optical depth and event rate assuming the wormhole lens to be bound or unbound to our Galaxy. We shall analyze the behavior of gutters appearing in the Paczy\'{n}skii light curves that typically characterize the backreacted charged wormholes. In the analysis, the effects of charges will be compared with those of the uncharged background spacetime, as well as with those of the Schwarzschild black hole.

The paper is organized as follows. In Sections 2 and 3, we shall briefly outline backreacted Kim and Lee wormholes with scalar and electric charges respectively but we mainly consider only the electrically charged wormhole. Sec.4 works out the light deflection angle correctly up to 4$^{\text{th}}$ PPN order and Sec.5 works out the lensing observables including Paczy\'{n}ski light curves. In Sec.6, we deal with the probabilistic characteristics of microlensing and conclude in Sec.7. We choose units such that $G=1$, $c=1$.

%%%%%%%%%%%%%%%%%%%%%%%%%%%%%%%%%%%%%%%%%
\section{Kim-Lee wormhole with scalar charge}\label{KL_scalar}
%%%%%%%%%%%%%%%%%%%%%%%%%%%%%%%%%%%%%%%%%
An outline of the backreacted scalar charge metric derived by Kim and Lee \cite{Kim:2001} is as follows: The Einstein equations for the background wormhole spacetime considered is 
\begin{equation}
G_{\mu \nu }^{(0)}=8\pi T_{\mu \nu }^{(0)}.
\end{equation}%
The left hand side, $G_{\mu \nu }^{(0)}$, is of the background wormhole geometry and the right hand side, $T_{\mu \nu }^{(0)}$, is the exotic matter threading that wormhole. If additional matter $T_{\mu \nu }^{(1)}$ is added to the right hand side and the corresponding backreaction $G_{\mu \nu}^{(1)} $ is added to the left hand side, then Einstein equations become 
\begin{equation}
G_{\mu \nu }^{(0)}+G_{\mu \nu }^{(1)}=8\pi \left[ T_{\mu \nu }^{(0)}+T_{\mu
\nu }^{(1)}\right].
\end{equation}

The authors consider a static Lorentzian wormhole with an additional minimally coupled massless scalar field stress-energy tensor given by 
\begin{eqnarray}
T_{\mu \nu }^{(1)} &=&\varphi _{;\mu }\varphi _{;\nu }-\frac{1}{2}g_{\mu \nu}g^{\rho \sigma }\varphi _{;\rho }\varphi _{;\sigma }, \\
\square \varphi &=&0
\end{eqnarray}%
and start with a background metric as a solution of Eq.(1), viz., 
\begin{equation}
d\tau ^{2}=-dt^{2}+\left( 1-\frac{b(r)}{r}\right) ^{-1}dr^{2}+r^{2}\left( d\theta ^{2}+\sin ^{2}{\theta }d\phi ^{2}\right) .
\end{equation}%
The backreacted metric due to scalar charge (denoted by the suffix "Sc") has been derived as the solution of Eq.(2), viz., 
\begin{equation}
d\tau _{\text{Sc}}^{2}=-dt^{2}+\left( 1-\frac{b(r)}{r}+\frac{\alpha }{r^{2}}\right) ^{-1}dr^{2}+r^{2}\left( d\theta ^{2}+\sin ^{2}{\theta }d\phi^{2}\right) ,
\end{equation}%
where the redshift function $\Phi =0$ and $\alpha $ plays the role of the scalar charge. For dimensional reasons, the function $b(r)$ is proposed in \cite{Kim:2001} as 
\begin{equation}
b(r)=b_{0}^{2\gamma /(2\gamma +1)}r^{1/(2\gamma +1)},
\end{equation}%
where $b_{0}$ is a constant, and the proper parameter $\gamma$ of the equation of state of matter threading the background wormhole (5) should be less than $-\frac{1}{2}$ so that the exponent of $r$ can be negative needed to satisfy the flareout condition. (Note: We have changed, in Eq.(7) and thereafter, their original notation $\beta$ to $\gamma$ to reserve the former for the source angle as is conventional in gravitational lensing).

The Kim-Lee scalar field $\varphi _{\text{KL}}$ as a solution of Eq.(4) is given by 
\begin{equation}
\varphi _{\text{KL}} =\varphi _{0}\left[ 1-\cos ^{-1}\left( \frac{b_{0}}{r}\right) \right] ,
\end{equation}%
where the free parameter $b_{0}$ has the dimension of length.

For the specific case $\gamma =-1$ studied by Kim and Lee \cite{Kim:2001}, one has $b(r) = b_{0}^{2}/r$. Then the metric (6) can be rewritten as 
\begin{equation}
d\tau _{\text{Sc}}^{2}=-dt^{2}+\left( 1-\frac{b_{0}^{2}-\alpha }{r^{2}}\right) ^{-1}dr^{2}+r^{2}\left( d\theta ^{2}+\sin ^{2}{\theta }d\phi^{2}\right) ,
\end{equation}%
where $r\in \lbrack 0,+\infty )$. We find that this metric formally resembles the massless Ellis-Bronnikov wormhole metric written in "standard coordinates" but with two differences introduced by the backreaction. First one is that the shape function of the background wormhole $b_{0}^{2}/r$ is shifted by an extra additive term $\frac{\alpha }{r}$ so that it is now given by $B_{\text{shape}}(r)=\left( b_{0}^{2}-\alpha \right) /r=a^{2}/r$ (say), where we have redefined $b_{0}^{2}-\alpha \equiv a^{2}$. The second one is that the Ellis-Bronnikov scalar field $\varphi_{\text{EB}}$ sourcing the wormhole (9) by a minimally coupled scalar field obtained from the massless case of the conformally rescaled Brans-Dicke theory \cite{Hayward:1996} is: 
\begin{equation}
\varphi_{\text{EB}} = \sqrt{2}\tan ^{-1}\left( \frac{\sqrt{r^{2}-a^{2}}}{a}\right) ,
\end{equation}%
which is different from $\varphi _{\text{KL}}$ of Eq.(8) though both functions tend to $\pi /2$ as $r\rightarrow +\infty $. The two functions, $\varphi _{\text{KL}}$ and $\varphi _{\text{EB}}$, generally differ but are identical, up to constant multiplicative factors, only when $\alpha =0$, or $a=b_{0}$, due to the identity cos$^{-1}\left( \frac{b_{0}}{r}\right) \equiv \tan ^{-1}\left( \frac{\sqrt{r^{2}-b_{0}^{2}}}{b_{0}}\right) $. Thus, we find that, for\ $\alpha \neq 0$ the solution (9) formally resembles, but is physically different from, that of the background wormhole as far as the source scalar field is concerned.

Following Morris and Thorne \cite{Morris:1988a}, the throat of the wormhole can be found using equation $B_{\text{shape}}(r_{\text{th}})=r_{\text{th}}$, so that we get 
\begin{equation}
r_{\text{th}}=\sqrt{b_{0}^{2}-\alpha },
\end{equation}%
which requires that $b_{0}^{2}>\alpha $ for a physically meaningful wormhole.

The asymptotic ADM mass is defined by 
\begin{equation}
M_{\text{ADM}}=\frac{1}{16\pi }\int \int_{S}\sum_{i,j=1}^{3}\left( \partial
_{j}g_{ij}-\partial _{i}g_{ii}\right) n^{i}dS,
\end{equation}%
where $S$ is a 2-surface enclosing the active region and $n^{i}$ denotes the unit outward normal. For the metric (9), one can find that $M=0$. This value coincides, as expected, with the asymptotic limit of the quasi-local Misner-Sharp (MS) mass within a ball of radius $r$ defined by \cite{Misner:1964, Hayward:1996}:%
\begin{equation}
M_{\text{MS}}=\frac{r}{2}\left( 1-g^{rr}\right) =\frac{b_{0}^{2}-\alpha }{2r}\rightarrow 0\text{ as}r\rightarrow \infty .
\end{equation}%
It should be noted that, by redefining $r^{2}=\ell ^{2}+a^{2}$, the metric (9) reduces to \cite{Nakajima:2012, Bhattacharya:2010}: 
\begin{eqnarray}
d\tau _{\text{EB}}^{2} &=&-dt^{2}+d\ell ^{2}+\left( \ell ^{2}+a^{2}\right)
\left( d\theta ^{2}+\sin ^{2}{\theta }d\phi ^{2}\right) , \\
\varphi _{\text{EB}} &=&\pm \sqrt{2}\tan ^{-1}\left( \frac{\ell }{a}\right)
\equiv \pm \sqrt{2}\left[ \pi /2-\tan ^{-1}\left( \frac{a}{\ell }\right) %
\right] \varpropto \mp \frac{a}{\ell },
\end{eqnarray}%
where $\ell \in (-\infty ,+\infty )$. This is just the familiar form of symmetric massless Ellis-Bronnikov wormhole with the scalar charge $\pm a$ residing on either side of the wormhole, one side attracting light towards the center while the other side repels it away from the center. Microlensing by this wormhole was studied in detail, e.g., in \cite{Abe:2010,Lukmanova:2018}. Therefore, we shall not pursue this case any further since it follows also as a corollary from the electrically charged wormhole when $Q=0$. The remainder of the paper will deal only with the effect of $Q\neq 0$ with $\Phi \neq 0$, which is the Kim-Lee wormhole with electric charge.

%%%%%%%%%%%%%%%%%%%%%%%%%%%%%%%%%%%%%%%%%
\section{Kim-Lee wormhole with electric charge}\label{KL_electr}
%%%%%%%%%%%%%%%%%%%%%%%%%%%%%%%%%%%%%%%%%
In this case, the additional stress-energy tensor for electromagnetic field is given by 
\begin{equation}
T_{\mu \nu }^{(1)}=\frac{1}{4\pi }\left( F_{\mu \lambda }F_{\nu }^{\lambda
}\right) -\frac{1}{4}g_{\mu \nu }F_{\lambda \sigma }F^{\lambda \sigma }
\end{equation}%
and the electric field for the charge $Q$ is 
\begin{equation}
E=\frac{Q}{r^{2}}\sqrt{g_{tt}g_{rr}}.
\end{equation}%
The general form of the backreacted wormhole metric due to electric charge (denoted by the suffix "El") is given by \cite{Kim:2001} 
\begin{equation}
d\tau _{\text{El}}^{2} = -\left( 1+\frac{Q^{2}}{r^{2}}\right) dt^{2}+\left( 1-\frac{b(r)}{r}+\frac{Q^{2}}{r^{2}}\right) ^{-1}dr^{2}+r^{2}\left( d\theta^{2}+\sin ^{2}\theta d\phi ^{2}\right),
\end{equation}%
where the redshift function $\Phi =\frac{1}{2}\ln \left( 1+\frac{Q^{2}}{r^{2}}\right) $, $Q$ is the electric charge and $b(r)$ is given, as before, by 
\begin{equation}
b(r)=b_{0}^{2\gamma /(2\gamma +1)}r^{1/(2\gamma +1)},
\end{equation}%
where $b_{0}$ is a constant. Once again, we will only consider the case $\gamma = - 1$ studied in \cite{Kim:2001}, which gives $b(r)=b_{0}^{2}/r$. Using these statements, metric (18) can be rewritten as 
\begin{equation}
d\tau _{\text{El}}^{2}=-\left( 1+\frac{Q^{2}}{r^{2}}\right) dt^{2}+\left( 1-%
\frac{b_{0}^{2}}{r^{2}}+\frac{Q^{2}}{r^{2}}\right) ^{-1}dr^{2}+r^{2}\left(
d\theta ^{2}+\sin ^{2}\theta d\phi ^{2}\right) .
\end{equation}%
The shape function is given by $B_{\text{shape}}(r)=\left(b_{0}^{2} - Q^{2}\right) /r$. The throat of the wormhole can be found using equation $B_{\text{shape}}(r_{\text{th}}) = r_{\text{th}}$ leading to $r_{\text{th}}$ and the quasi-local Misner-Sharp (MS) mass as 
\begin{eqnarray}
r_{\text{th}} &=&\sqrt{b_{0}^{2}-Q^{2}}. \\
M_{\text{MS}} &=&\left. \frac{b_{0}^{2}-Q^{2}}{2r}\right\vert _{r=r_{\text{th%
}}}=r_{\text{th}}/2.
\end{eqnarray}%
In order for the throat radius to exist, it is necessary to have $Q^{2}<b_{0}^{2}$. The wormhole has no real non-zero photon sphere radius since it requires $2Q^{2}+r_{ph}^{2}=0$ unless $Q=0$, in which case the spacetime becomes the background massless Ellis-Bronnikov wormhole.

It can be directly verified that the ADM mass of the electrically charged wormholes (20) is $M_{\text{ADM}} = 0$ consisent with $M_{\text{MS}} \rightarrow 0$ as $r\rightarrow \infty $. It means that we can consider this solution as a massless charged wormhole. Further in this work, we will compare the profiles of microlensing of charged wormholes with the profiles of an electrically neutral massless Ellis-Bronnikov wormhole and a Schwarzschild black hole.

%%%%%%%%%%%%%%%%%%%%%%%%%%%%%%%%%%%%%%%%%
\section{Light deflection}\label{light_defl}
%%%%%%%%%%%%%%%%%%%%%%%%%%%%%%%%%%%%%%%%%
We employ Keeton-Petters method \cite{Keeton:2005} to calculate the light deflection angle and lensing observables. Start with the general form of metric in standard coordinates given by 
\begin{equation}
d\tau ^{2}=-A(r)dt^{2}+B(r)dr^{2}+r^{2}\left( d\theta ^{2}+\sin ^{2}{\theta }d\phi ^{2}\right) ,
\end{equation}%
and PPN expand the asymptotically flat metric functions in powers of $b_{0}/r $ as 
\begin{eqnarray}
A(r) &=&1-2a_{1}\left( \frac{b_{0}}{r}\right) +2a_{2}\left( \frac{b_{0}}{r}\right) ^{2}-2a_{3}\left( \frac{b_{0}}{r}\right) ^{3}+2a_{4}\left( \frac{b_{0}}{r}\right) ^{4}+... \\
B(r) &=&1+2b_{1}\left( \frac{b_{0}}{r}\right) +4b_{2}\left( \frac{b_{0}}{r}\right) ^{2}+8b_{3}\left( \frac{b_{0}}{r}\right) ^{3}+16b_{4}\left( \frac{b_{0}}{r}\right) ^{4}+...
\end{eqnarray}%
where $a_{1}$, $a_{2}$, $a_{3}$, $a_{4}$, $b_{1}$, $b_{2}$, $b_{3}$ and $b_{4}$ are the constant coefficients. Then the deflection angle up to 4$^{\text{th}}$ PPN order is given by 
\begin{equation}
\widehat{\alpha }(b)=A_{1}\left( \frac{b_{0}}{b}\right) +A_{2}\left( \frac{b_{0}}{b}\right) ^{2}+A_{3}\left( \frac{b_{0}}{b}\right) ^{3}+A_{4}\left(\frac{b_{0}}{b}\right) ^{4}+...,
\end{equation}%
where $b$ is the impact parameter and 
\begin{eqnarray}
A_{1} &=&2(a_{1}+b_{1}), \\
A_{2} &=&\pi \left( 2a_{1}^{2}+a_{1}b_{1}-a_{2}-\frac{b_{1}^{2}}{4}+b_{2}\right) , \\
A_{3} &=&\frac{2}{3}\left(35a_{1}^{3}+15a_{1}^{2}b_{1}-30a_{1}a_{2}-3a_{1}b_{1}^{2}+12a_{1}b_{2}-6a_{2}b_{1}+6a_{3}+b_{1}^{3}-4b_{1}b_{2}+8b_{3}\right) , \\
A_{4} &=&\pi \left[ 30a_{1}^{4}+12a_{1}^{3}b_{1}-\frac{9}{4}a_{1}^{2}\left(
16a_{2}+b_{1}^{2}-4b_{2}\right) +a_{1}\left( 9a_{3}-9a_{2}b_{1}+\frac{%
3b_{1}^{3}}{4}-3b_{1}b_{1}+6b_{3}\right) \right.  \notag \\
&&\left. +\frac{3}{64}\left( 96a_{2}^{2}+16a_{2}\left( b_{1}^{2}-4b_{2}\right)
+32a_{3}b_{1}-32a_{4}-5b_{1}^{4}+24b_{1}^{2}b_{2}-32b_{1}b_{3}-16b_{2}^{2}+64b_{4}\right) \right] .
\end{eqnarray}%
Note that $A_{4}$ is our new contribution in this paper over the known Keeton-Petters \cite{Keeton:2005} coefficients $A_{1}$, $A_{2}$, $A_{3}$.

The expansion of metric functions in (23) in powers of $b_{0}/r$ has the
form 
\begin{eqnarray}
A(r) &=&1+\frac{Q^{2}}{b_{0}^{2}}\left( \frac{b_{0}}{r}\right) ^{2}, \\
B(r) &=&1+\left( 1-\frac{Q^{2}}{b_{0}^{2}}\right) \left( \frac{b_{0}}{r}%
\right) ^{2}+\left( 1-\frac{Q^{2}}{b_{0}^{2}}\right) ^{2}\left( \frac{b_{0}}{%
r}\right) ^{4}+...
\end{eqnarray}%
So we can finally obtain the PPN coefficients 
\begin{eqnarray}
a_{1} &=&a_{3}=a_{4}=0,\quad a_{2}=\frac{Q^{2}}{2b_{0}^{2}}, \\
b_{1} &=&b_{3}=0,\quad b_{2}=\frac{1}{4}\left( 1-\frac{Q^{2}}{b_{0}^{2}}%
\right) ,\quad b_{4}=\frac{1}{16}\left( 1-\frac{Q^{2}}{b_{0}^{2}}\right)^{2}.
\end{eqnarray}%
which yield 
\begin{eqnarray}
A_{1} &=&0, \\
A_{2} &=&\frac{\pi }{4}\left( 1-\frac{3Q^{2}}{b_{0}^{2}}\right) , \\
A_{3} &=&0, \\
A_{4} &=&\frac{3\pi }{64}\left( 3-\frac{14Q^{2}}{b_{0}^{2}}+\frac{35Q^{4}}{b_{0}^{4}}\right) .
\end{eqnarray}

The bending angle can then be expressed as a series in powers of the single quantity $\left( b_{0}/b<<1\right) $: 
\begin{equation}
\widehat{\alpha }_{\text{El}}(b)=\frac{\pi }{4}\left( 1-\frac{3Q^{2}}{b_{0}^{2}}\right) \left( \frac{b_{0}}{b}\right) ^{2}+\frac{3\pi }{64}\left(3-\frac{14Q^{2}}{b_{0}^{2}}+\frac{35Q^{4}}{b_{0}^{4}}\right) \left( \frac{b_{0}}{b}\right) ^{4}+....
\end{equation}%
When $Q=0$, we exactly recover the background massless Ellis-Bronnikov wormhole for which \cite{Nakajima:2012, Bhattacharya:2010}\footnote{%
Gao et al \cite{Gao:2023} obtained a negative sign, probably a typo, before the 4$^{\text{th}}$ order term $\frac{9\pi }{64}\left( \frac{b_{0}}{b}\right) ^{4}$ in Eq.(40) in disagreement with the correct positive sign obtained in \cite{Nakajima:2012, Bhattacharya:2010}.}%
\begin{equation}
\widehat{\alpha }_{\text{EB}}(b)=\frac{\pi }{4}\left( \frac{b_{0}}{b}\right)^{2}+\frac{9\pi }{64}\left( \frac{b_{0}}{b}\right) ^{4}+...
\end{equation}%
From the equation (39), we can conclude that the presence of electric charge $Q\neq 0$ decreases the deflection $\widehat{\alpha }_{\text{El}}(b)$ from the background value (40). We also find an interesting result that the leading order deflection angle is zero at $Q^{2}=b_{0}^{2}/3$ and at $Q^{2}<b_{0}^{2}/3$, the deflection is positive, while at $Q^{2}>b_{0}^{2}/3$ it is negative! The positivity of throat constrains $Q^{2}<b_{0}^{2}$, whereas positivity of deflection angle imposes a stronger constraint $Q^{2}<b_{0}^{2}/3$. Its impact on lensing observables will be explored soon.

\begin{figure}[!ht]
  \centerline{\includegraphics[scale=0.75]{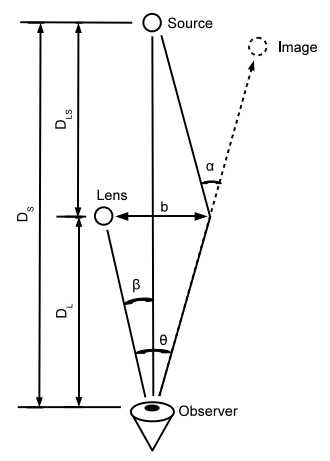}}
  \caption{Lens geometry, where $D_{L}$ is the distance from the observer to the lens, $D_{S}$ is the distance from the observer to the source, $D_{LS}$ is the distance from the lens to the source, $b$ is the impact parameter of the light, $\beta $ is the angle between lens and source, $\theta $ is the angle between the image and lens, and $\alpha $ is the bending angle between the image and source. Figure taken from \cite{Abe:2010}.}
\end{figure}

%%%%%%%%%%%%%%%%%%%%%%%%%%%%%%%%%%%%%%%%%
\section{Lensing observables}\label{lens_obs}
%%%%%%%%%%%%%%%%%%%%%%%%%%%%%%%%%%%%%%%%%
We first need to define the Einstein radius $R_{E}$ and the angular radius $\theta_{E}$ of the Einstein ring at the lens plane. In the weak field limit, $b$ is approximately equal to the closest approach distance $r$ and for the charged wormhole (20), they are related by \cite{Keeton:2005} 
\begin{equation}
\frac{1}{b_{0}^{2}}=\frac{A(r)}{r^{2}}=\frac{1}{r^{2}}\left( 1+\frac{Q^{2}}{%
r^{2}}\right) \simeq \frac{1}{r^{2}}\text{ as }r\rightarrow \infty .
\end{equation}
The deflection angle to leading order can be expressed as, from Eq.(39), 
\begin{equation}
\widehat{\alpha }_{\text{El}}(b)=\frac{\pi \left( b_{0}^{2}-3Q^{2}\right) }{4b^{2}} + O\left( \frac{1}{b}\right) ^{4}.
\end{equation}%
The source angle $\beta $ between the optical axis (the line joining the wormhole lens and the observer) and the line joining the observer and the source star can be written from the elementary lensing geometry \cite{Virbhadra:2000, Virbhadra:2002, Keeton:2005} for light traversing the lens on two opposite sides as (see Fig.1 for lensing geometry): 
\begin{equation}
\beta =\frac{b}{D_{L}}\pm \frac{D_{LS}}{D_{S}}\widehat{\alpha }_{\text{El}}(b),
\end{equation}%
where the Euclidean distances from the observer to the lens is $D_{L}$ and to the source is $D_{S}$, the distance from the lens to the source is $D_{LS}=D_{S}-D_{L}$, respectively and $b$ is the impact parameter. As evident in Eq.(42), the deflection angle of the electrically charged wormhole starts from $O\left(1 / r^{2}\right) $. This is due to the massless nature of the considered wormhole and indicates the possibility of observational discrimination from the gravitational lensing effect of black holes which, to leading order, is of $O\left( 1/r\right) $. Following Abe \cite{Abe:2010}, we can introduce a proper coordinate $\ell \in (-\infty ,+\infty )$ in the deflection angle Eq.(39) by the relation $b^{2}=\ell ^{2}+(b_{0}^{2}-3Q^{2})$, which yields $b\rightarrow \ell $ as $\ell \rightarrow \infty $. We thus obtain from Eqs.(43) 
\begin{equation}
\beta =\frac{\ell }{D_{L}}-\frac{\pi \left( b_{0}^{2}-3Q^{2}\right) }{4\ell^{2}}\frac{D_{LS}}{D_{S}}\hspace{1cm}(\ell >0),
\end{equation}%
while the source angle on the opposite side is: 
\begin{equation}
\beta =\frac{\ell }{D_{L}}+\frac{\pi \left( b_{0}^{2}-3Q^{2}\right) }{4\ell^{2}}\frac{D_{LS}}{D_{S}}\hspace{1cm}(\ell <0).
\end{equation}

\begin{figure}[!ht]
  \centerline{\includegraphics[scale=0.4]{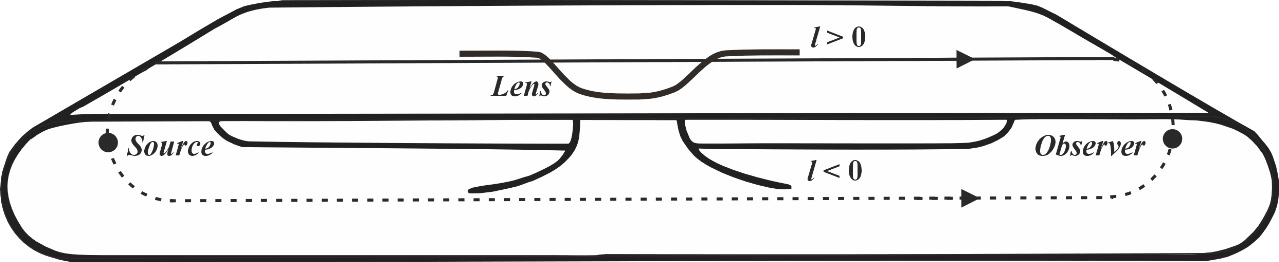}}
  \caption{Equatorial slices identified at infinity in the Morris-Thorne embedding diagram (see Ref.\cite{Morris:1988a}).}
\end{figure}

At this point, it is necessary to clarify why $\pm$ signs appear in Eq.(43), or more explicitly, in the two image angle Eqs.(44) and (45) in the regimes $\ell >0$ and $\ell <0$ respectively. Then it might seem that light has to inevitably pass through the throat while going from one regime to the other so that the weak lensing approximation breaks down
\footnote{%
We sincerely thank an anonymous reviewer for raising this conceptual point.}. Such a breakdown can be avoided by appropriately choosing the null geodesics. We illustrate the scenario in terms of the simplest massless Ellis-Bronnikov (EB) wormhole Eq.(14) that has a throat defined by the minimum radius, $\ell _{\text{th}}=0$. Let us now introduce in Eq.(14) the "standard" radial coordinate $r$ defined by $r=\sqrt{\ell ^{2}+a^{2}}$. The resultant metric then has a throat defined by $r_{\text{th}}=a$ such that $r\in (a, \infty)$. The weak field light deflection to leading order is given by $\widehat{\alpha }_{\text{EB}}(r)\simeq \frac{\pi a^{2}}{4r^{2}}$. The deflection preserves the same form to the leading order, viz., we have $\widehat{\alpha }_{\text{EB}}(\ell )\simeq \frac{\pi a^{2}}{4\ell ^{2}}$, as can be seen by direct integration (see for details \cite{Bhattacharya:2010}) or simply by using the conversion $r=\sqrt{\ell ^{2}+a^{2}}\rightarrow \ell $ as $\ell \rightarrow \infty $ in $\widehat{\alpha }_{\text{EB}}(r)$.
 
The interesting thing is that the throat $\ell _{\text{th}}$ is a photon sphere, i.e., $\ell _{\text{ps}}=0$ \cite{Nakajima:2012, Bhattacharya:2010}, which translates to $r_{\text{ps}}=a$. Therefore, a null geodesic passing vertically through the wormhole throat would be captured forever at the throat so that it never emerges on the other side and lensing cannot occur at all in this case! The only possibility for lensing to occur then is to consider null geodesics that do not encounter the photon sphere. To achieve this scenario, we first identify at infinity the equatorial slices of the two asymptotically flat regions of the wormhole metric (14) and assume that the source and observer are located on the opposite sides of the lens (wormhole throat). Next, we assume that two null geodesics emerge from the source, pass horizontally along the identified slices above and below the lens as shown in the Morris-Thorne embedding diagram (see Fig.2). Light rays undergo deflection $\widehat{\alpha }_{\text{EB}}(\ell )\simeq \frac{\pi a^{2}}{4\ell ^{2}}$ at the mouths tagged by the $\pm$ signs in Eqs.(44), (45) for $\beta $ to indicate the horizontal light motions in the $\ell > 0$ and $\ell < 0$ regimes that finally reach the observer at infinity on the other side of the lens.

The electrically charged Kim-Lee wormhole, Eq.(20), is fundamentally no different from the chargeless Ellis-Bronnikov wormhole: Both are massless, symmetric wormholes with the first order deflection by the Kim-Lee wormhole being $\widehat{\alpha }_{\text{El}}(\ell )\simeq \frac{\pi(b_{0}^{2}-3Q^{2})}{4\ell ^{2}}$ yielding eventually Eqs.(44) and (45) for source angles $\beta$. One then obtains Eqs.(77) and (78) with the corresponding image angles $\theta > 0$ and $\theta < 0$.

In our work, we further assumed the null geodesics to pass far from the lens (in the weak field region) and legitimately used the Keeton-Petters approach. However, when the light rays pass arbitrarily close by the photon sphere, the PPN method no longer works \cite{Tsukamoto:2017} and strong field lensing methods need to be applied. In the rest of the paper beginning from Eq.(47) below, we shall resume the original standard coordinate of the Kim-Lee wormhole, Eq.(20), with $r_{\text{th}}\neq 0$, for which the Keeton-Petters method can be applied to analyse the lensing observables.

\begin{table}
\begin{tabular}{|c|c|c|c||c|c|c|c|}
\hline
$r_{\text{th}}$ [km] & $Q/r_{\text{th}}$ & $R_{E}$ [km] & $\theta_{E}$ [mas]
& $r_{\text{th}}$ [km] & $Q/r_{\text{th}}$ & $R_{E}$ [km] & $\theta_{E}$
[mas] \\ \hline
$10^{0}$ & $0$ & $3.6457\times 10^{5}$ & $0.0006$ & $10^{6}$ & $0$ & $%
3.6458\times 10^{9}$ & $6.0940$ \\ 
$10^{0}$ & $0.3$ & $3.4124\times 10^{5}$ & $0.0006$ & $10^{6}$ & $0.3$ & $%
3.4124\times 10^{9}$ & $5.7039$ \\ 
$10^{0}$ & $0.5$ & $2.8937\times 10^{5}$ & $0.0005$ & $10^{6}$ & $0.5$ & $%
2.8937\times 10^{9}$ & $4.8368$ \\ 
$10^{0}$ & $0.7$ & $9.8962\times 10^{4}$ & $0.0004$ & $10^{6}$ & $0.7$ & $%
9.8962\times 10^{8}$ & $1.6542$ \\ \hline
$10^{1}$ & $0$ & $1.6922\times 10^{6}$ & $0.0028$ & $10^{7}$ & $0$ & $%
1.6922\times 10^{10}$ & $28.2858$ \\ 
$10^{1}$ & $0.3$ & $1.5839\times 10^{6}$ & $0.0026$ & $10^{7}$ & $0.3$ & $%
1.5839\times 10^{10}$ & $26.4752$ \\ 
$10^{1}$ & $0.5$ & $1.3431\times 10^{6}$ & $0.0022$ & $10^{7}$ & $0.5$ & $%
1.3431\times 10^{10}$ & $22.4505$ \\ 
$10^{1}$ & $0.7$ & $4.5934\times 10^{5}$ & $0.0008$ & $10^{7}$ & $0.7$ & $%
4.5934\times 10^{9}$ & $7.6779$ \\ \hline
$10^{2}$ & $0$ & $7.8546\times 10^{6}$ & $0.0131$ & $10^{8}$ & $0$ & $%
7.8546\times 10^{10}$ & $131.2911$ \\ 
$10^{2}$ & $0.3$ & $7.3518\times 10^{6}$ & $0.0123$ & $10^{8}$ & $0.3$ & $%
7.3518\times 10^{10}$ & $122.8872$ \\ 
$10^{2}$ & $0.5$ & $6.2342\times 10^{6}$ & $0.0104$ & $10^{8}$ & $0.5$ & $%
6.2342\times 10^{10}$ & $104.2058$ \\ 
$10^{2}$ & $0.7$ & $2.1321\times 10^{6}$ & $0.0036$ & $10^{8}$ & $0.7$ & $%
2.1321\times 10^{10}$ & $35.6379$ \\ \hline
$10^{3}$ & $0$ & $3.6457\times 10^{7}$ & $0.0609$ & $10^{9}$ & $0$ & $%
3.6458\times 10^{11}$ & $609.3995$ \\ 
$10^{3}$ & $0.3$ & $3.4124\times 10^{7}$ & $0.0570$ & $10^{9}$ & $0.3$ & $%
3.4124\times 10^{11}$ & $570.3919$ \\ 
$10^{3}$ & $0.5$ & $2.8937\times 10^{7}$ & $0.0484$ & $10^{9}$ & $0.5$ & $%
2.8937\times 10^{11}$ & $483.6807$ \\ 
$10^{3}$ & $0.7$ & $9.8962\times 10^{6}$ & $0.0165$ & $10^{9}$ & $0.7$ & $%
9.8962\times 10^{10}$ & $165.4165$ \\ \hline
$10^{4}$ & $0$ & $1.6922\times 10^{8}$ & $0.2829$ & $10^{10}$ & $0$ & $%
1.6922\times 10^{12}$ & $2828.5817$ \\ 
$10^{4}$ & $0.3$ & $1.5839\times 10^{8}$ & $0.2647$ & $10^{10}$ & $0.3$ & $%
1.5839\times 10^{12}$ & $2647.5247$ \\ 
$10^{4}$ & $0.5$ & $1.3431\times 10^{8}$ & $0.2245$ & $10^{10}$ & $0.5$ & $%
1.3431\times 10^{12}$ & $2245.0468$ \\ 
$10^{4}$ & $0.7$ & $4.5934\times 10^{7}$ & $0.0768$ & $10^{10}$ & $0.7$ & $%
4.5934\times 10^{11}$ & $767.7952$ \\ \hline
$10^{5}$ & $0$ & $7.8546\times 10^{8}$ & $1.3129$ & $10^{11}$ & $0$ & $%
7.8546\times 10^{12}$ & $13129.1133$ \\ 
$10^{5}$ & $0.3$ & $7.3518\times 10^{8}$ & $1.2289$ & $10^{11}$ & $0.3$ & $%
7.3518\times 10^{12}$ & $12288.7209$ \\ 
$10^{5}$ & $0.5$ & $6.2342\times 10^{8}$ & $1.0421$ & $10^{11}$ & $0.5$ & $%
6.2342\times 10^{12}$ & $10420.5841$ \\ 
$10^{5}$ & $0.7$ & $2.1321\times 10^{8}$ & $0.3564$ & $10^{11}$ & $0.7$ & $%
2.1321\times 10^{12}$ & $3563.7896$ \\ \hline
\end{tabular}
\caption{Einstein radius $R_{E}$ and its angular radius $\theta _{E}$ in milliarcsec (mas) for the lens configuration when the source ($S$) star is in the Galactic Bulge and the wormhole lens ($L$) is at some intermediate distance ($D_{L}=4$ kpc and $D_{S}=8$ kpc assumed). The charge $Q$ is scaled by $r_{\text{th}}$ under the restriction that $\frac{Q}{r_{\text{th}}}<\frac{1}{\sqrt{2}}=0.707$ so that $R_{E}$ is real [see Eq.(47)].}
\end{table}

\begin{table}
\begin{tabular}{|c|c|c|c||c|c|c|c|}
\hline
$r_{\text{th}}$ [km] & $Q/r_{\text{th}}$ & $R_{E}$ [km] & $\theta_{E}$ [mas]
& $r_{\text{th}}$ [km] & $Q/r_{\text{th}}$ & $R_{E}$ [km] & $\theta_{E}$
[mas] \\ \hline
$10^{0}$ & $0$ & $6.7156\times 10^{5}$ & $0.0002$ & $10^{6}$ & $0$ & $%
6.7156\times 10^{9}$ & $1.7960$ \\ 
$10^{0}$ & $0.3$ & $6.2857\times 10^{5}$ & $0.0002$ & $10^{6}$ & $0.3$ & $%
6.2857\times 10^{9}$ & $1.6811$ \\ 
$10^{0}$ & $0.5$ & $5.3302\times 10^{5}$ & $0.0001$ & $10^{6}$ & $0.5$ & $%
5.3302\times 10^{9}$ & $1.4255$ \\ 
$10^{0}$ & $0.7$ & $1.8229\times 10^{5}$ & $0.0001$ & $10^{6}$ & $0.7$ & $%
1.8229\times 10^{9}$ & $0.4875$ \\ \hline
$10^{1}$ & $0$ & $3.1171\times 10^{6}$ & $0.0008$ & $10^{7}$ & $0$ & $%
3.1171\times 10^{10}$ & $8.3365$ \\ 
$10^{1}$ & $0.3$ & $2.9176\times 10^{6}$ & $0.0008$ & $10^{7}$ & $0.3$ & $%
2.9176\times 10^{10}$ & $7.8028$ \\ 
$10^{1}$ & $0.5$ & $2.4740\times 10^{6}$ & $0.0007$ & $10^{7}$ & $0.5$ & $%
2.4740\times 10^{10}$ & $6.6167$ \\ 
$10^{1}$ & $0.7$ & $8.4611\times 10^{5}$ & $0.0002$ & $10^{7}$ & $0.7$ & $%
8.4611\times 10^{9}$ & $2.2629$ \\ \hline
$10^{2}$ & $0$ & $1.4468\times 10^{7}$ & $0.0039$ & $10^{8}$ & $0$ & $%
1.4468\times 10^{11}$ & $38.6945$ \\ 
$10^{2}$ & $0.3$ & $1.3542\times 10^{7}$ & $0.0036$ & $10^{8}$ & $0.3$ & $%
1.3542\times 10^{11}$ & $36.2176$ \\ 
$10^{2}$ & $0.5$ & $1.1483\times 10^{7}$ & $0.0031$ & $10^{8}$ & $0.5$ & $%
1.1483\times 10^{11}$ & $30.7118$ \\ 
$10^{2}$ & $0.7$ & $3.9273\times 10^{6}$ & $0.0010$ & $10^{8}$ & $0.7$ & $%
3.9273\times 10^{10}$ & $10.5033$ \\ \hline
$10^{3}$ & $0$ & $6.7156\times 10^{7}$ & $0.0180$ & $10^{9}$ & $0$ & $%
6.7156\times 10^{11}$ & $179.6037$ \\ 
$10^{3}$ & $0.3$ & $6.2857\times 10^{7}$ & $0.0168$ & $10^{9}$ & $0.3$ & $%
6.2857\times 10^{11}$ & $168.1073$ \\ 
$10^{3}$ & $0.5$ & $5.3302\times 10^{7}$ & $0.0142$ & $10^{9}$ & $0.5$ & $%
5.3302\times 10^{11}$ & $142.5516$ \\ 
$10^{3}$ & $0.7$ & $1.8229\times 10^{7}$ & $0.0049$ & $10^{9}$ & $0.7$ & $%
1.8229\times 10^{11}$ & $48.7520$ \\ \hline
$10^{4}$ & $0$ & $3.1171\times 10^{8}$ & $0.0834$ & $10^{10}$ & $0$ & $%
3.1171\times 10^{12}$ & $833.6467$ \\ 
$10^{4}$ & $0.3$ & $2.9176\times 10^{8}$ & $0.0780$ & $10^{10}$ & $0.3$ & $%
2.9176\times 10^{12}$ & $780.2851$ \\ 
$10^{4}$ & $0.5$ & $2.4740\times 10^{8}$ & $0.0662$ & $10^{10}$ & $0.5$ & $%
2.4740\times 10^{12}$ & $661.6658$ \\ 
$10^{4}$ & $0.7$ & $8.4611\times 10^{7}$ & $0.0226$ & $10^{10}$ & $0.7$ & $%
8.4611\times 10^{11}$ & $226.2865$ \\ \hline
$10^{5}$ & $0$ & $1.4468\times 10^{9}$ & $0.3869$ & $10^{11}$ & $0$ & $%
1.4468\times 10^{13}$ & $3869.4453$ \\ 
$10^{5}$ & $0.3$ & $1.3542\times 10^{9}$ & $0.3622$ & $10^{11}$ & $0.3$ & $%
1.3542\times 10^{13}$ & $3621.7628$ \\ 
$10^{5}$ & $0.5$ & $1.1483\times 10^{9}$ & $0.3071$ & $10^{11}$ & $0.5$ & $%
1.1483\times 10^{13}$ & $3071.1808$ \\ 
$10^{5}$ & $0.7$ & $3.9273\times 10^{8}$ & $0.1050$ & $10^{11}$ & $0.7$ & $%
3.9273\times 10^{12}$ & $1050.3291$ \\ \hline
\end{tabular}
\caption{Einstein radius $R_{E}$ and its angular radius $\theta _{E}$ in milliarcsec (mas) for lens configuration when source star is in the LMC ($D_{L}=25$ kpc and $D_{S}=50$ kpc assumed) lensed by electrically charged wormhole. The charge $Q$ is scaled by $r_{\text{th}}$ under the restriction that $\frac{Q}{r_{\text{th}}}<\frac{1}{\sqrt{2}}=0.707$ so that $R_{E}$ is real [see Eq.(47)].}
\end{table}

If the source and lens are exactly aligned along the line of sight (i.e., $\beta =0$), the image is expected to be circular (Einstein ring). The Einstein radius $r=R_{E}$, which is defined as the radius of the circular image at the lens plane, is obtained by solving the cubic Eq.(44) with $\beta =0$ yielding a positive root 
\begin{eqnarray}
R_{E} &=& \left[\frac{\pi }{4}\frac{D_{L}D_{LS}\left(b_{0}^{2}-3Q^{2}\right) }{D_{S}}\right] ^{1/3} \\
&=&\left[ \frac{\pi }{4}\frac{D_{L}D_{LS}\left( 1-\frac{2Q^{2}}{r_{\text{th}}^{2}}\right) r_{\text{th}}^{2}}{D_{S}}\right] ^{1/3},
\end{eqnarray}%
where we have used Eq.(21) for the throat radius $r_{\text{th}}$. From Eq.(46) it follows that the Einstein radius will take real values only under the stronger constraint that $Q^{2}<b_{0}^{2}/3$, which translates into a canonical constraint 
\begin{equation}
\frac{Q}{r_{\text{th}}}<\frac{1}{\sqrt{2}}\simeq 0.707.
\end{equation}

Using these statements, the image positions can be calculated from 
\begin{equation}
\beta =\theta -\frac{\theta _{E}^{3}}{\theta ^{2}}\hspace{1cm}(\theta > 0)
\end{equation}%
and 
\begin{equation}
\beta =\theta +\frac{\theta _{E}^{3}}{\theta ^{2}}\hspace{1cm}(\theta < 0),
\end{equation}%
where $\theta =b/D_{L}\thickapprox r/D_{L}$ is the angle between the image and lens and 
\begin{equation}
\theta _{E}=R_{E}/D_{L}
\end{equation}%
is the angular Einstein radius. Following Keeton-Petters \cite{Keeton:2005}, we expand the angular position of the image as 
\begin{equation}
\theta =\theta _{0}+\theta _{1}\varepsilon +\theta _{2}\varepsilon
^{2}+...\;,
\end{equation}%
where $\varepsilon $ is a dimensionless small parameter defined by 
\begin{equation}
\varepsilon =\frac{\theta _{E}D_{S}}{4D_{LS}}=\left( \frac{D_{S}}{%
4D_{L}D_{LS}}\right) R_{E}.
\end{equation}%
For the Galactic microlensing under consideration, the distances, where $D_{L}$, $D_{S}$, $D_{LS}\equiv D_{S}-D_{L}$ and $R_{E}$ are such that $\varepsilon \approx 10^{-7}$ (Table 1), hence the expansion (52) is well justified. Indeed, from Table 1, it can be seen that the Einstein angle $\theta _{E}\approx 10^{-5\prime \prime }$ to $10^{-4\prime \prime}$, well justifying Galactic microlensing of the Bulge stars. Measuring this tiny angle is beyond the capability of current technology. However, lensing by wormholes can still be observable by measuring the \textit{change} in brightness in the images of the stars due to the relative motion of the star and the wormhole lens. Light curves will be analyzed in the sequel to expose this feature of lensing.

%%%%%%%%%%%%%%%%%%%%%%%%%
\subsection{Image position}
\label{IP}
%%%%%%%%%%%%%%%%%%%%%%%%%

\begin{figure}
  \centerline{\includegraphics[scale=0.38]{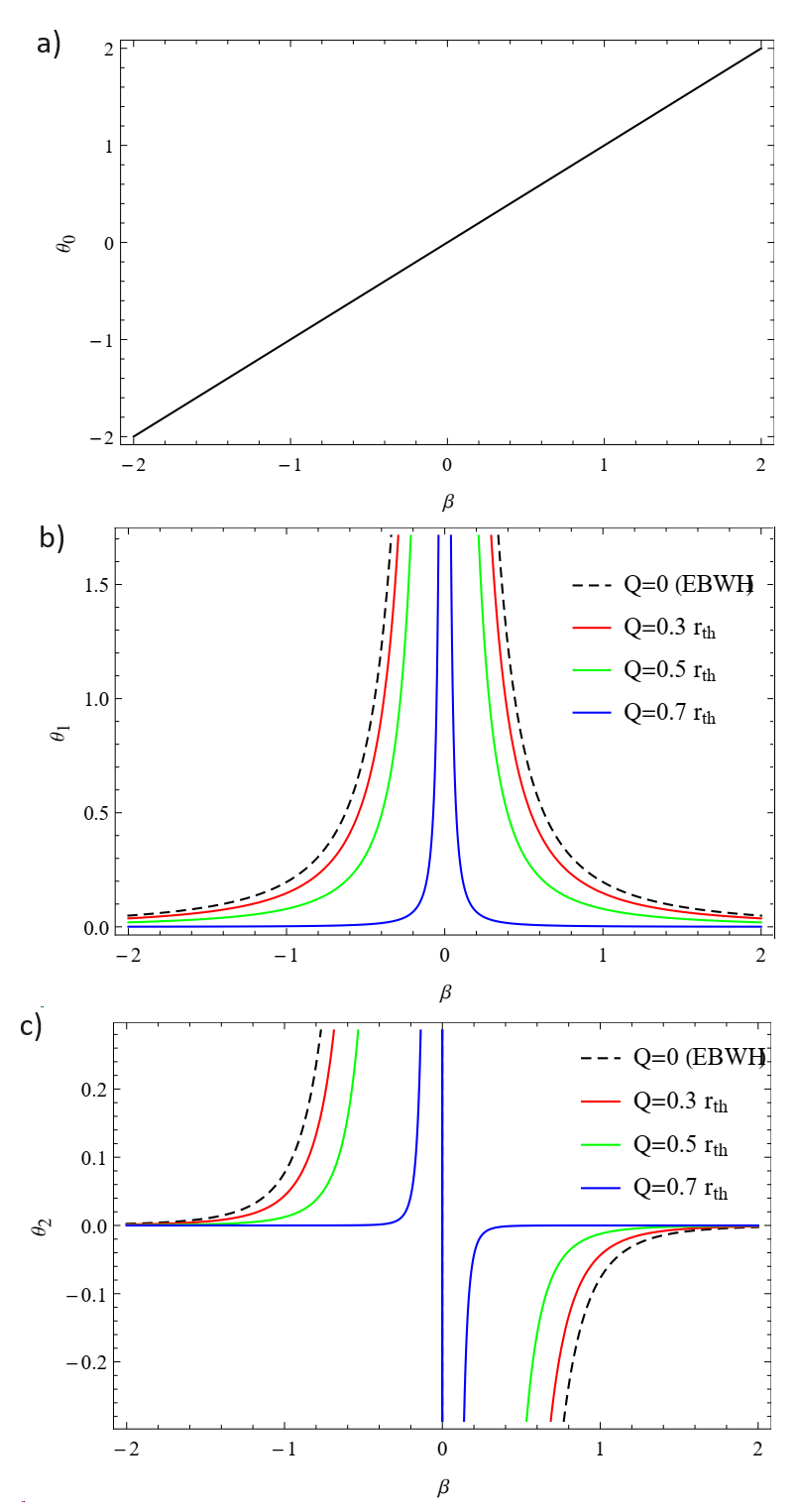}}
  \caption{Angular image position of the source lenced by Kim-Lee wormhole as a function of the angular source position $\beta $ ($\beta >0$ corresponds to the positive-parity image while $\beta <0$ corresponds to the negative-parity image) and $Q$. Fig.3a shows the zeroth-order image position $\theta _{0}$, which is independent on charge $Q$. Fig.3b shows the first-order correction term of image position $\theta _{1}$ for $Q=0$ (massless Ellis-Bronnikov wormhole indicated by EBWH) -- dashed black line, $Q=0.3r_{\text{th}}$ -- solid red line, $Q = 0.5 r_{\text{th}}$ -- solid green line, and $Q=0.7r_{\text{th}}$ -- solid blue line. Fig.3c shows the second-order correction term of image position $\theta _{2}$ for values of charge $Q$.}
\label{IP}
\end{figure}

Here, $\theta _{0}$ in the expansion (52) is the zeroth order image position in the weak deflection limit, while $\theta _{1}$ and $\theta _{2}$ are the first- and second order PPN correction terms to the image position $\theta _{0}$. Following Keeton-Petters method \cite{Keeton:2005}, the angles of image positions are set to be positive. It means that the position of the source is positive if the image is on the same side of the lens as the source (positive parity image), while it is negative if the image is on the opposite side (negative parity image). The image position $\theta _{0}$ and corrections $\theta _{1}$ and $\theta _{2}$ to the image position of stars are completely determined by the coefficients $A_{i}$, with $A_{1}=0$, and work out to%
\begin{eqnarray}
\theta _{0}^{\text{(El)}} &=&\beta , \\
\theta _{1}^{\text{(El)}} &=&\frac{\pi }{16\beta ^{2}}\left( 1-\frac{3Q^{2}}{b_{0}^{2}}\right)  \nonumber \\
&=&\frac{\pi }{16\beta ^{2}}\left[ 1-\frac{3Q^{2}}{r_{\text{th}}}^{2}\left(1+\frac{Q^{2}}{r_{\text{th}}^{2}} \right) ^{-1}\right] , \\
\theta _{2}^{\text{(El)}} &=&-\frac{\pi ^{2}}{128\beta ^{5}}\left( 1-\frac{3Q^{2}}{b_{0}^{2}}\right) ^{2}  \nonumber \\
&=&-\frac{\pi ^{2}}{128\beta ^{5}}\left[ 1-\frac{3Q^{2}}{r_{\text{th}}^{2}}\left( 1+\frac{Q^{2}}{r_{\text{th}}}^{2}\right) ^{-1}\right] .
\end{eqnarray}

Figs.3a-c show the corrections to the zeroth image position $\theta_{0}^{\text{(El)}}$ of the source lensed by electrically charged wormhole. The quantity $\theta_{0}^{\text{(El)}}$ does not depend on electric charge $Q$ as shown in Fig.3a but remarkably coincides with the source position $\beta$. This peculiarity arises because $A_{1}=2(a_{1}+b_{1})=0$ with the leading order light deflection appearing only at the second order $\varpropto \left(\frac{b_{0}}{b}\right)^{2}$ probably characteristic of massless wormholes. Images then appear only at the corrective angles $\theta_{i}$. The positive parity (same side of the lens) and negative parity (opposite side of the lens) image positions correspond respectivly to $\beta > 0$ and $\beta < 0$ respectively. The case $Q=0$ corresponds to the background Ellis-Bronnikov wormhole. For $Q\neq 0$, the corrections to image positions lensed by the background wormhole is always greater in magnitude than those by the electrically charged wormhole, as evident from Eqs.(55,56). It can be seen from the Figs.3b,c that for the canonical constraint $\frac{Q}{r_{\text{th}}}<\frac{1}{\sqrt{2}}$, the first correction to the image position lies in the positive region, while with increase of electric charge the corrections decrease. The extreme case $Q=r_{\text{th}}/\sqrt{2}$ gives $\theta _{1}^{\text{(El)}} = \theta _{2}^{\text{(El)}}=0$, which means that there is no correction to the zeroth image position, that is, image cannot be distinguished from the source. If such a phenomenon is really observed, that is, there appears no multiple images, then the lens could possibly be an extremely charged one. Fig.3c plots the second order corrections to the image position. These features indicate theoretical distinctions between two massless wormholes, the background and the electrically charged backreacted ones. The background observables are not accessible to observers since the charge $Q$ of an astrophysical lens cannot be switched on and off at will to measure the distinctions.

%%%%%%%%%%%%%%%%%%%%%%%%%
\subsection{Magnification}
\label{Magnification}
%%%%%%%%%%%%%%%%%%%%%%%%%

\begin{figure}
  \centerline{\includegraphics[scale=0.5]{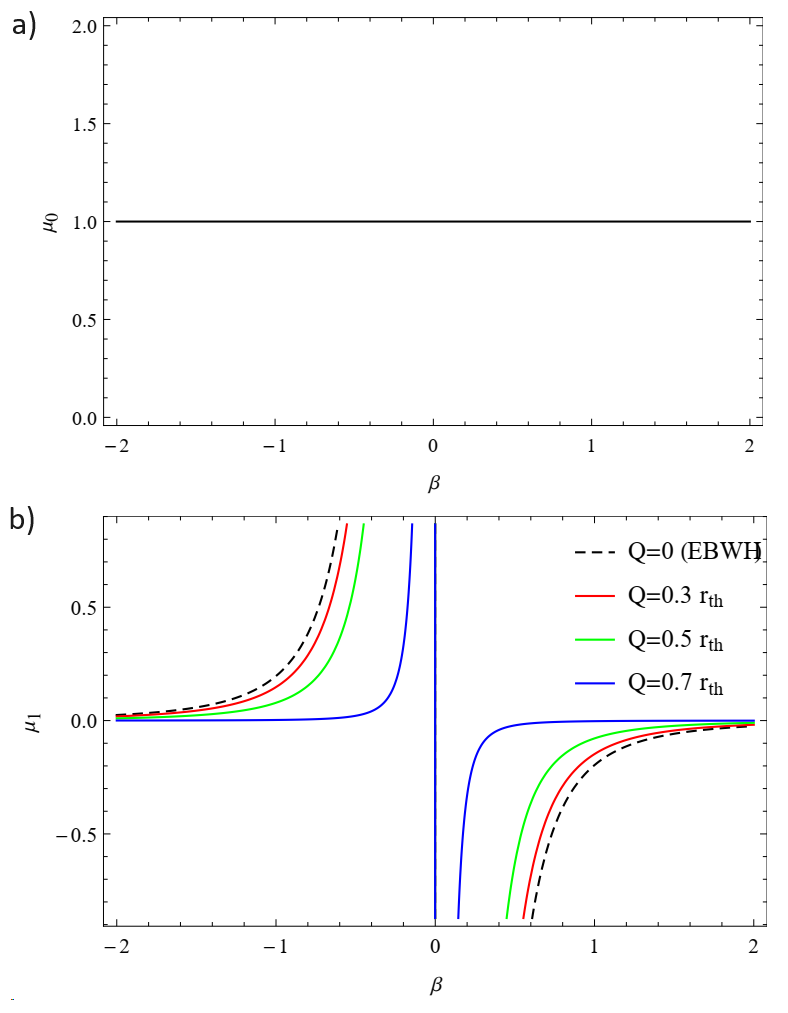}}
  \caption{ Image magnification for Kim-Lee wormhole as a function of the angular source position $\beta $ and charge $Q$. Fig.4a shows the zeroth-order image magnification $\mu _{0}$, which is independent on charge $Q$. Fig.4b shows the first-order correction term of image magnification $\mu _{1}$ for $Q=0$ (massless Ellis-Bronnikov wormhole indicated by EBWH) -- dashed black line, $Q = 0.3r_{\text{th}}$ -- solid red line, $Q=0.5r_{\text{th}}$ -- solid green line, and $Q=0.7r_{\text{th}}$ -- solid blue line.}
\end{figure}

The magnification of an image is defined by the ratio between the solid angles of the image and the source, defined by 
\begin{equation}
\mu \left( \theta \right) =\left[ \frac{\sin \beta \mathfrak{(\theta )}}{%
\sin \theta }\frac{d\beta \mathfrak{(\theta )}}{d\theta }\right] ^{-1}.
\end{equation}%
Applying the same algorithm as for image position, we assume the series expansion of magnification in the form 
\begin{equation}
\mu =\mu _{0}+\mu _{1}\varepsilon +\mu _{2}\varepsilon ^{2}+...\;,
\end{equation}%
where $\mu _{0}$ is the zeroth order magnification, $\mu_{1}$ and $\mu_{2}$ are the first and second order PPN correction terms, which work out to 
\begin{eqnarray}
\mu _{0}^{\text{(El)}} &=&1, \\
\mu _{1}^{\text{(El)}} &=&-\frac{\pi }{16\beta ^{2}}\left[ 1-\frac{3Q^{2}}{r_{\text{th}}^{2}}\left( 1+\frac{Q^{2}}{r_{\text{th}}^{2}}\right) ^{-1}\right] , \\
\mu _{2}^{\text{(El)}} &=&0.
\end{eqnarray}

Figs.4a,b show the zeroth and higher order of magnification of the image. Fig.4a shows that the zeroth order does not depend on electric charge and is equal to a constant value. Accordingly, all influence can be traced in the first order. In Fig.4b, it can be seen that the largest magnification of the image corresponds to the case when the charge $Q=0$, i.e., the case of the Ellis-Bronnikov wormhole. Increasing the parameter $Q$ leads to a reduction in the resulting magnification. Equation (61) shows that the second order correction to image magnification is identically zero.

%%%%%%%%%%%%%%%%%%%%%%%%%
\subsection{Total magnification and Centroid}
\label{tot_magn} 
%%%%%%%%%%%%%%%%%%%%%%%%%

\begin{figure}
  \centerline{\includegraphics[scale=0.5]{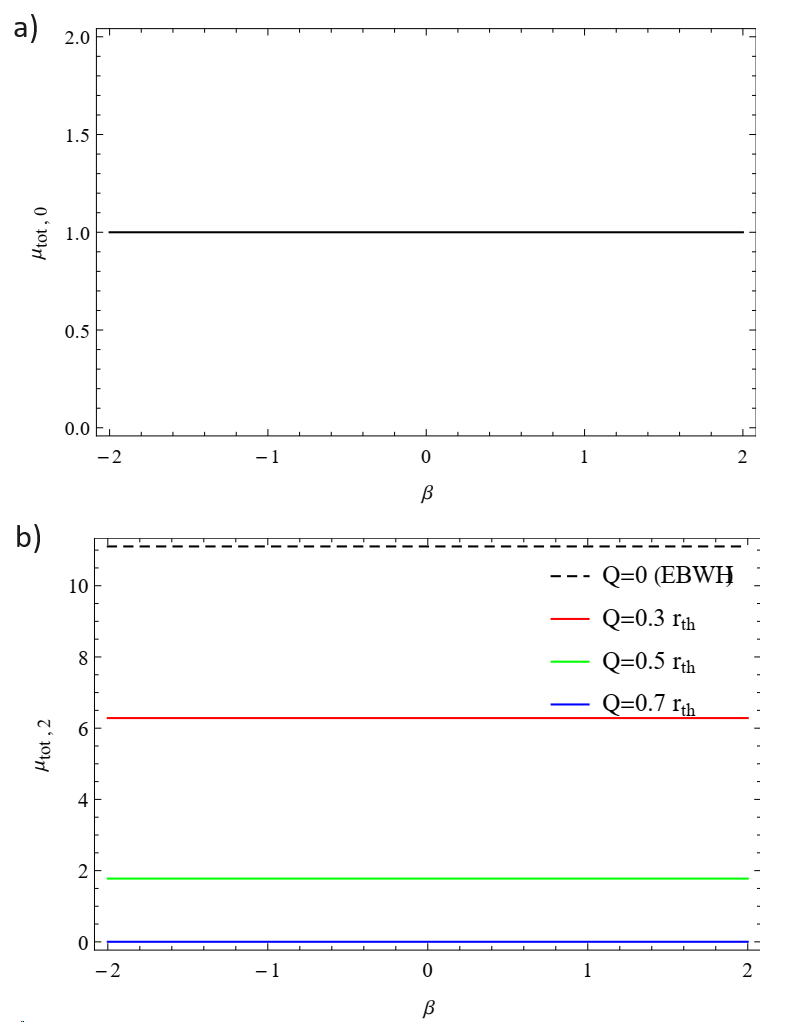}}
  \caption{The total magnification of images for Kim-Lee wormhole as a function of the angular source position $\beta $ and charge $Q$. Fig.5a shows the zeroth-order total magnification $\mu _{\text{tot},0}$, which does not depend on either $\beta $ or $Q$. Since the first-order correction of the total magnification $\mu _{\text{tot},1}$ is zero, we plotted only the second-order correction $\mu _{\text{tot},2}$ in Fig.5b for $Q=0$ (massless Ellis-Bronnikov wormhole indicated by EBWH) -- dashed black line, $Q=0.3r_{\text{th}}$ -- solid red line, $Q=0.5r_{\text{th}}$ -- solid green line, and $Q=0.7r_{\text{th}}$ -- solid blue line. The second-order correction does not depend on angular source position $\beta$, but depends only on the charge $Q$.}
\end{figure}

\begin{figure}[!ht]
  \centerline{\includegraphics[scale=0.5]{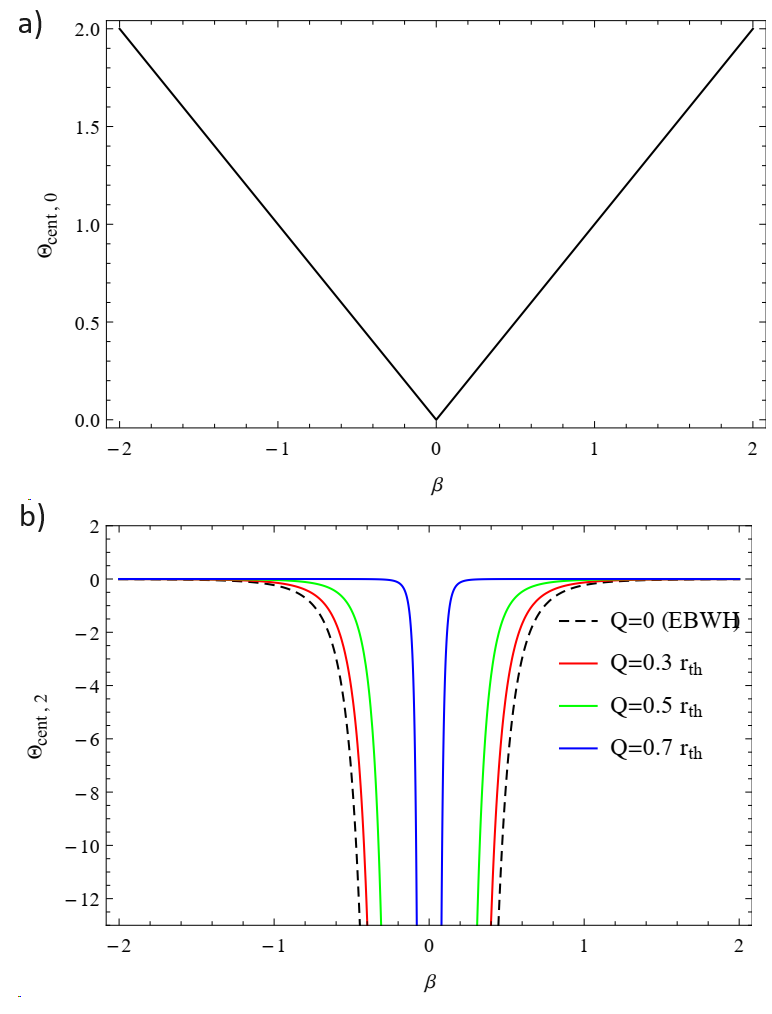}}
  \caption{The magnification-weighted centroid position for Kim-Lee wormhole as a function of the angular source position $\beta $ and charge $Q$. Fig.6a shows the zeroth-order magnification-weighted centroid position $\Theta _{\text{cent},0}$, which is independent on charge $Q$. Since the first-order correction of the total magnification $\Theta _{\text{cent},1}$ is zero, we plotted only the second-order correction $\Theta _{\text{cent},2}$ in Fig.6b for $Q=0$ (massless Ellis-Bronnikov wormhole indicated by EBWH) -- dashed black line, $Q=0.3r_{\text{th}}$ -- solid red line, $Q=0.5r_{\text{th}}$ -- solid green line, and $Q=0.7r_{\text{th}}$ -- solid blue line.}
\end{figure}

In the case when the individual images are too close together and cannot be resolved, it is useful to define total magnification $\mu_{\text{tot}}$ as a sum of magnification of positive and negative parity images $\mu ^{+}$ and $\mu ^{-}$ corresponding to $\beta > 0$ and $\beta < 0$ respectively, as 
\begin{equation}
\mu _{\text{tot}} = |\mu ^{+}|+|\mu ^{-}| = \mu _{\text{tot}, 0} + \mu _{\text{tot}, 1} \varepsilon +\mu _{\text{tot}, 2} \varepsilon^{2} + ...\;,
\end{equation}%
where for electrically charged wormhole are 
\begin{eqnarray}
\mu _{\text{tot}, 0}^{\text{(El)}} &=&1, \\
\mu _{\text{tot}, 1}^{\text{(El)}} &=&0, \\
\mu _{\text{tot}, 2}^{\text{(El)}} &=&\frac{9\pi ^{2}}{8}\left( 1-\frac{3Q^{2}}{b_{0}^{2}}\right) ^{2}  \notag \\
&=&\frac{9\pi ^{2}}{8}\left[ 1-\frac{3Q^{2}}{r_{\text{th}}^{2}}\left( 1+ \frac{Q^{2}}{r_{\text{th}}^{2}}\right)^{-1}\right] ^{2}.
\end{eqnarray}

Figs.5a,b show the zero and second orders the total image magnification of source lenced by electrically charged Kim-Lee wormhole. Fig.5a shows that the zero order does not depend on electric charge, and is equal to a constant value, as in the previous case. Accordingly, all influence can be traced only to the second order, because the first order correction equals zero.

Fig.5b plots the second order correction for the overall magnification. It can be seen from the figure that an increase in the parameter $Q$ leads to a decrease in the total magnification and reaches its lowest value at $Q=b_{0}/\sqrt{3}$. With a further decrease in $Q$, the overall image magnification increases.

The center of light or in short the centroid $\Theta _{\text{cent}}$ of the images is simply the magnification-weighted sum of the image positions and its expansion is 
\begin{eqnarray}
\Theta _{\text{cent}} &=&\frac{\theta ^{+}|\mu ^{+}|+\theta ^{-}|\mu ^{-}|}{%
|\mu ^{+}|-|\mu ^{-}|}  \notag \\
&=&\Theta _{\text{cent}, 0} + \Theta _{\text{cent}, 1}\varepsilon +\Theta _{%
\text{cent}, 2} \varepsilon ^{2}+...\;.
\end{eqnarray}%
where for electrically charged wormhole are 
\begin{eqnarray}
\Theta _{\text{cent}, 0}^{\text{(El)}} &=&|\beta |, \\
\Theta _{\text{cent}, 1}^{\text{(El)}} &=&0, \\
\Theta _{\text{cent}, 2}^{\text{(El)}} &=&-\frac{3\pi ^{2}}{128|\beta ^{5}|}%
\left( 1-\frac{3Q^{2}}{b_{0}^{2}}\right) ^{2} \\
&=&-\frac{3\pi ^{2}}{128|\beta ^{5}|}\left[ 1-\frac{3Q^{2}}{r_{\text{th}}^{2}%
}\left( 1+\frac{Q^{2}}{r_{\text{th}}^{2}}\right) ^{-1}\right] ^{2}.
\end{eqnarray}

Figs.6a,b show the zeroth and second orders of the centroid by the wormhole with an electric charge. Fig.6a shows that the zero order does not depend on electric charge. Accordingly, the entire influence of the electric charge can be traced in the second order, because the first order correction is $0$.

%%%%%%%%%%%%%%%%%%%%%%%%%
\subsection{Differential time delay}
\label{TD}
%%%%%%%%%%%%%%%%%%%%%%%%% 

The absolute value of the time delay from a single lensed image is never accessible in the weak field lensing \cite{Keeton:2005}. The practically observable diagnostic is the differential time delay between the positive- and negative-parity images $\hat{\tau}_{+}$ and $\hat{\tau}_{-}$ as 
\begin{equation}
\Delta \hat{\tau}=\hat{\tau}_{-}-\hat{\tau}_{+}.
\end{equation}%
The differential time delay $\Delta \hat{\tau}$ is the delay in arrival times at the observer can also be PPN expanded in term of $\varepsilon$ as
\begin{equation}
\Delta \hat{\tau}=\Delta \hat{\tau}_{0}\ +\ \Delta \hat{\tau}%
_{1}\,\varepsilon \ +\ O\left( {\varepsilon }\right) ^{2},
\end{equation}%
where 
\begin{eqnarray}
\Delta \hat{\tau}_{0} &=&\frac{1}{2}\,|\beta |\sqrt{A_{1}+\beta ^{2}}+\frac{%
A_{1}}{4}\ \ln \left( \frac{\sqrt{A_{1}+\beta ^{2}}+\beta }{\sqrt{%
A_{1}+\beta ^{2}}-\beta }\right) \,, \\
\Delta \hat{\tau}_{1} &=&\frac{A_{2}}{A_{1}}\ |\beta |\,,
\end{eqnarray}%
which yields, in the limit $A_{1}\rightarrow 0$, 
\begin{eqnarray}
\Delta \hat{\tau}_{0} &=&\frac{1}{2}\beta |\beta |, \\
\Delta \hat{\tau}_{1} &\rightarrow &\infty .
\end{eqnarray}%
The result of $\Delta \hat{\tau}_{1}$ is same as for massless Ellis-Bronnikov wormhole as in \cite{Lukmanova:2018}. Remarkably, time delay up to $O\left( {\varepsilon }\right) ^{2}$ is independent of the charge of wormhole. Note further that, if $A_{2}=0$ but $\beta \neq 0$, then from Eq.(74), $\Delta \hat{\tau}_{1} = 0$ and from Eq.(73), it follows that the measurement of $\Delta \hat{\tau}_{0}$ would make it possible to determine location of the angular position $\beta $ of the source. On the other hand, if $A_{2}\neq 0$ but $\beta =0$ or the source, lens, and observer are aligned then one has an Einstein ring from where light reaches the observer exactly at the same time so that there is no time delay, a fact symbolized by $\Delta \hat{\tau}_{0}=0.$ Since there are no individual images now, the corrections to individual images also lose their meaning, a symptom of which is the appearance of divergences in Eq.(76).

%%%%%%%%%%%%%%%%%%%%%%%%%
\subsection{Paczy\'{n}ski light curves}
\label{LC}
%%%%%%%%%%%%%%%%%%%%%%%%%

So far we had used Keeton-Petters method \cite{Keeton:2005} for obtaining the lensing observables but for calculating the light curves, we will use Abe's \cite{Abe:2010} approach. Using reduced parameters $\hat{\beta}=\beta /\theta _{E}$ and $\hat{\theta} = \theta /\theta _{E}$, Eqs. (49) and (50) become simple cubic equations: 
\begin{equation}
\hat{\theta}^{3}-\hat{\beta}\hat{\theta}^{2}-1=0\hspace{1cm}(\hat{\theta}>0)
\end{equation}%
and 
\begin{equation}
\hat{\theta}^{3}-\hat{\beta}\hat{\theta}^{2}+1=0\hspace{1cm}(\hat{\theta}<0).
\end{equation}
The above cubic equations can in principle have at most three real solutions of $\theta$. Each non-degenerate real solution represents an image (see, e.g., discussions in \cite{Asada:2017}), so there can be at most three images by the Kim and Lee charged wormholes. The number of images plays an important role in microlensing effect of a wormhole \cite{Safonova:2002}. On the other hand, Eq.(62) states that the total magnification $\mu _{+} + \mu _{-}$ is the summation due only to two images indicating an apparent discrepancy between the total magnification and the number of images. However, we shall first illustrate by a simple example that there are indeed two images only, one inside and the other outside the Einstein ring on the lens plane.

Consider the cubic Eq.(77) with $\widehat{\beta }=2=\frac{\beta }{\theta _{E}}$ so that $\theta _{E} = \frac{\beta }{2}$. Now assume $\theta _{E}=1$ (say) so that $\beta =2$. Then the cubic equation becomes\footnote{%
This illustrative example of $\widehat{\beta }=2$ was discussed by an anonymous reviewer.} 
\begin{equation*}
\theta ^{3}-4\theta ^{2}-1=0,\text{ }\theta >0.
\end{equation*}%
This yields two complex conjugate roots and one real root $\theta =4.06>\theta _{E}$, that is the image appears outside the Einstein ring. Similarly, the other cubic Eq.(78)%
\begin{equation*}
\theta ^{3}-4\theta ^{2}+1=0,\text{ }\theta <0
\end{equation*}%
has three real roots $\theta =-0.47$, $0.53$, $3.93$. From these roots, the condition $\theta <0$ picks out only $\theta =-0.47$, which shows that the image appears inside the Einstein ring $\theta _{E}=1$. This situation is clearly portrayed also in \cite{Abe:2010}. The total magnification is thus due only to two images $\mu_{+}$ and $\mu _{-}$.

Next, we will generically solve the cubic Eqs.(77), (78) for $\hat{\theta}$ using Vi\`{e}te's trigonometric method. The solution for $\hat{\beta}<(27/4)^{1/3}$ is 
\begin{equation}
\hat{\theta}_{1}=\frac{\hat{\beta}}{3}+\frac{2\hat{\beta}}{3}\sin \left[ \frac{\pi }{6}+\frac{1}{3}\cos ^{-1}\left( -\frac{2\hat{\beta}^{3}+27}{2\hat{\beta}^{3}}\right) \right] ,
\end{equation}%
and for $\hat{\beta}\geq (27/4)^{1/3}$, it is 
\begin{equation}
\hat{\theta}_{1}=\frac{\hat{\beta}}{3}+\frac{2\hat{\beta}}{3}\text{sgn}\left( -2\hat{\beta}^{3}-27\right) \cosh \left[ \frac{1}{3}\cosh ^{-1}\left(\frac{|-2\hat{\beta}^{3}-27|}{2\hat{\beta}^{3}}\right) \right] .
\end{equation}%
Similarly, for $\hat{\beta}<(27/4)^{1/3}$, one has 
\begin{equation}
\hat{\theta}_{2}=\frac{\hat{\beta}}{3}-\frac{2\hat{\beta}}{3}\text{sgn}%
\left( -2\hat{\beta}^{3}+27\right) \cosh \left[ \frac{1}{3}\cosh ^{-1}\left( 
\frac{|-2\hat{\beta}^{3}+27|}{2\hat{\beta}^{3}}\right) \right] ,
\end{equation}%
and for $\hat{\beta}\geq (27/4)^{1/3},$ one has 
\begin{equation}
\hat{\theta}_{2}=\frac{\hat{\beta}}{3}-\frac{2\hat{\beta}}{3}\cos \left[ 
\frac{1}{3}\cos ^{-1}\left( -\frac{2\hat{\beta}^{3}+27}{2\hat{\beta}^{3}}%
\right) \right] .
\end{equation}

The above roots differ in form from those of the massless Ellis-Bronnikov wormhole, but graphically they coincide with those derived by Abe \cite{Abe:2010}. However, the effect of charge $Q$ is not \textit{directly} manifest in the expression for these roots but they appear in the light curves of the charged wormhole via $\hat{\beta},\hat{\theta}$, which are functions of $\theta _{E}$ that contains $Q$ through $\theta _{E}$ $\left( \equiv R_{E}/D_{L}\right) $, where $R_{E}$ is given by Eq.(47). The light curves are defined by the magnification $\mu (t)$ given by Paczy\'{n}ski formula \cite{Paczynski:1986}
\begin{eqnarray}
\mu (t) &=&\left\vert \frac{\hat{\theta}_{1}}{\hat{\beta}}\frac{d\hat{\theta}%
_{1}}{d\hat{\beta}}\right\vert +\left\vert \frac{\hat{\theta}_{2}}{\hat{\beta%
}}\frac{d\hat{\theta}_{2}}{d\hat{\beta}}\right\vert \\
&=&\left\vert \left[ \left( 1-\frac{1}{\hat{\theta}_{1}^{3}}\right) \left( 1+%
\frac{2}{\hat{\theta}_{1}^{3}}\right) \right] ^{-1}\right\vert +\left\vert %
\left[ \left( 1+\frac{1}{\hat{\theta}_{2}^{3}}\right) \left( 1-\frac{2}{\hat{%
\theta}_{2}^{3}}\right) \right] ^{-1}\right\vert ,
\end{eqnarray}%
with time dependence of $\hat{\beta}$ as 
\begin{equation}
\hat{\beta}(t)=\sqrt{\hat{\beta}_{0}+\left( t-t_{0}\right) ^{2}/t_{E}^{2}},
\end{equation}%
where $\hat{\beta}_{0}$ is the impact parameter of the source trajectory, $t_{0}$ is the time of closest approach and $t_{E}$ is the Einstein radius crossing time given by 
\begin{equation}
t_{E}=R_{E}/v_{T},
\end{equation}%
where $v_{T}$ is the transverse velocity of the source relative to the lens and observer.

\begin{table}
\begin{tabular}{|c|c|c|c||c|c|c|c|}
\hline
$r_{\text{th}}$ [km] & $Q/r_{\text{th}}$ & $t_{E}$ [day] & $t_{E}$ [day] & $%
r_{\text{th}}$ [km] & $Q/r_{\text{th}}$ & $t_{E}$ [day] & $t_{E}$ [day] \\ 
&  & Bound & Unbound &  &  & Bound & Unbound \\ \hline
$10^{0}$ & $0$ & $0.0192$ & $0.0008$ & $10^{6}$ & $0$ & $191.8029$ & $8.4393$
\\ 
$10^{0}$ & $0.3$ & $0.0179$ & $0.0008$ & $10^{6}$ & $0.3$ & $179.5257$ & $%
7.8991$ \\ 
$10^{0}$ & $0.5$ & $0.0152$ & $0.0007$ & $10^{6}$ & $0.5$ & $152.2341$ & $%
6.6983$ \\ 
$10^{0}$ & $0.7$ & $0.0052$ & $0.0002$ & $10^{6}$ & $0.7$ & $52.0633$ & $%
2.2908$ \\ \hline
$10^{1}$ & $0$ & $0.0890$ & $0.0039$ & $10^{7}$ & $0$ & $890.2704$ & $%
39.1719 $ \\ 
$10^{1}$ & $0.3$ & $0.0833$ & $0.0037$ & $10^{7}$ & $0.3$ & $833.2843$ & $%
36.6645$ \\ 
$10^{1}$ & $0.5$ & $0.0707$ & $0.0031$ & $10^{7}$ & $0.5$ & $706.6081$ & $%
31.0908$ \\ 
$10^{1}$ & $0.7$ & $0.0242$ & $0.0011$ & $10^{7}$ & $0.7$ & $241.6566$ & $%
10.6329$ \\ \hline
$10^{2}$ & $0$ & $0.4132$ & $0.0182$ & $10^{8}$ & $0$ & $4132.2690$ & $%
181.8198$ \\ 
$10^{2}$ & $0.3$ & $0.3867$ & $0.0170$ & $10^{8}$ & $0.3$ & $3867.7632$ & $%
170.1816$ \\ 
$10^{2}$ & $0.5$ & $0.3280$ & $0.0144$ & $10^{8}$ & $0.5$ & $3279.7841$ & $%
144.3105$ \\ 
$10^{2}$ & $0.7$ & $0.1121$ & $0.0049$ & $10^{8}$ & $0.7$ & $1121.6704$ & $%
49.3535$ \\ \hline
$10^{3}$ & $0$ & $1.9180$ & $0.0844$ & $10^{9}$ & $0$ & $19180.2939$ & $%
843.9329$ \\ 
$10^{3}$ & $0.3$ & $1.7953$ & $0.0790$ & $10^{9}$ & $0.3$ & $17952.5664$ & $%
789.9129$ \\ 
$10^{3}$ & $0.5$ & $1.5223$ & $0.0670$ & $10^{9}$ & $0.5$ & $15223.4094$ & $%
669.8300$ \\ 
$10^{3}$ & $0.7$ & $0.5206$ & $0.0229$ & $10^{9}$ & $0.7$ & $5206.3328$ & $%
229.0786$ \\ \hline
$10^{4}$ & $0$ & $8.9027$ & $0.3917$ & $10^{10}$ & $0$ & $89027.0382$ & $%
3917.1897$ \\ 
$10^{4}$ & $0.3$ & $8.3328$ & $0.3666$ & $10^{10}$ & $0.3$ & $83328.4319$ & $%
3666.4510$ \\ 
$10^{4}$ & $0.5$ & $7.0661$ & $0.3109$ & $10^{10}$ & $0.5$ & $70660.8070$ & $%
3109.0755$ \\ 
$10^{4}$ & $0.7$ & $2.4166$ & $0.1063$ & $10^{10}$ & $0.7$ & $24165.6561$ & $%
1063.2889$ \\ \hline
$10^{5}$ & $0$ & $41.3227$ & $1.8182$ & $10^{11}$ & $0$ & $413226.9063$ & $%
18181.9839$ \\ 
$10^{5}$ & $0.3$ & $38.6776$ & $1.7018$ & $10^{11}$ & $0.3$ & $386776.3191$
& $17018.1580$ \\ 
$10^{5}$ & $0.5$ & $32.7978$ & $1.4431$ & $10^{11}$ & $0.5$ & $327978.4129$
& $14431.0502$ \\ 
$10^{5}$ & $0.7$ & $11.2167$ & $0.4935$ & $10^{11}$ & $0.7$ & $112167.0394$
& $4935.3497$ \\ \hline
\end{tabular}
\caption{The Einstein radius crossing time $t_{E}$ by the source in the Bulge ($%
D_{L}=4$ kpc and $D_{S}=8$ kpc assumed). When the wormhole lens is bound to
the Galaxy (the transverse velocity $v_{T}$ of the lens relative
to the source star and observer, assumed to be $v_{T}=220$ km/s) and when
the lens is unbound, the velocity is assumed to be $v_{T}=5000$ km/s). The
charge $Q$ is scaled by $r_{\text{th}}$ under the restriction that $\frac{Q%
}{r_{\text{th}}}<\frac{1}{\sqrt{2}}=0.707$ so that $R_{E}$ is real [see
Eq.(47)]}
\end{table}

\begin{table}
\begin{tabular}{|c|c|c|c||c|c|c|c|}
\hline
$r_{\text{th}}$ [km] & $Q/r_{\text{th}}$ & $t_{E}$ [day] & $t_{E}$ [day] & $%
r_{\text{th}}$ [km] & $Q/r_{\text{th}}$ & $t_{E}$ [day] & $t_{E}$ [day] \\ 
&  & Bound & Unbound &  &  & Bound & Unbound \\ \hline
$10^{0}$ & $0$ & $0.0353$ & $0.0015$ & $10^{6}$ & $0$ & $353.3040$ & $%
15.5454 $ \\ 
$10^{0}$ & $0.3$ & $0.0331$ & $0.0014$ & $10^{6}$ & $0.3$ & $330.6891$ & $%
14.5503$ \\ 
$10^{0}$ & $0.5$ & $0.0280$ & $0.0012$ & $10^{6}$ & $0.5$ & $280.4176$ & $%
12.3384$ \\ 
$10^{0}$ & $0.7$ & $0.0096$ & $0.0004$ & $10^{6}$ & $0.7$ & $95.9015$ & $%
4.2197$ \\ \hline
$10^{1}$ & $0$ & $0.1640$ & $0.0072$ & $10^{7}$ & $0$ & $1639.8921$ & $%
72.1553$ \\ 
$10^{1}$ & $0.3$ & $0.1535$ & $0.0067$ & $10^{7}$ & $0.3$ & $1534.9228$ & $%
67.5366$ \\ 
$10^{1}$ & $0.5$ & $0.1301$ & $0.0057$ & $10^{7}$ & $0.5$ & $1301.5832$ & $%
57.2697$ \\ 
$10^{1}$ & $0.7$ & $0.0445$ & $0.0020$ & $10^{7}$ & $0.7$ & $445.1352$ & $%
19.5859$ \\ \hline
$10^{2}$ & $0$ & $0.7612$ & $0.0335$ & $10^{8}$ & $0$ & $7611.7047$ & $%
334.9150$ \\ 
$10^{2}$ & $0.3$ & $0.7124$ & $0.0313$ & $10^{8}$ & $0.3$ & $7124.4807$ & $%
313.4771$ \\ 
$10^{2}$ & $0.5$ & $0.6041$ & $0.0266$ & $10^{8}$ & $0.5$ & $6041.4140$ & $%
265.8222$ \\ 
$10^{2}$ & $0.7$ & $0.2066$ & $0.0091$ & $10^{8}$ & $0.7$ & $2066.1345$ & $%
90.9099$ \\ \hline
$10^{3}$ & $0$ & $3.5330$ & $0.1554$ & $10^{9}$ & $0$ & $35330.4035$ & $%
1554.5377$ \\ 
$10^{3}$ & $0.3$ & $3.3069$ & $0.1455$ & $10^{9}$ & $0.3$ & $33068.9101$ & $%
1455.0320$ \\ 
$10^{3}$ & $0.5$ & $2.8042$ & $0.1234$ & $10^{9}$ & $0.5$ & $28041.7598$ & $%
1233.8374$ \\ 
$10^{3}$ & $0.7$ & $0.9590$ & $0.0422$ & $10^{9}$ & $0.7$ & $9590.1470$ & $%
421.9665$ \\ \hline
$10^{4}$ & $0$ & $16.3989$ & $0.7215$ & $10^{10}$ & $0$ & $163989.2064$ & $%
7215.5251$ \\ 
$10^{4}$ & $0.3$ & $15.3492$ & $0.6754$ & $10^{10}$ & $0.3$ & $153492.2839$
& $6753.6605$ \\ 
$10^{4}$ & $0.5$ & $13.0158$ & $0.5727$ & $10^{10}$ & $0.5$ & $130158.3194$
& $5726.9660$ \\ 
$10^{4}$ & $0.7$ & $4.4513$ & $0.1959$ & $10^{10}$ & $0.7$ & $44513.5191$ & $%
1958.5948$ \\ \hline
$10^{5}$ & $0$ & $76.1170$ & $3.3491$ & $10^{11}$ & $0$ & $761170.4694$ & $%
33491.5006$ \\ 
$10^{5}$ & $0.3$ & $71.2448$ & $3.1348$ & $10^{11}$ & $0.3$ & $712448.0712$
& $31347.7151$ \\ 
$10^{5}$ & $0.5$ & $60.4141$ & $2.6582$ & $10^{11}$ & $0.5$ & $604141.4020$
& $26582.2217$ \\ 
$10^{5}$ & $0.7$ & $20.6613$ & $0.9091$ & $10^{11}$ & $0.7$ & $206613.4531$
& $9090.9919$ \\ \hline
\end{tabular}
\caption{The Einstein radius crossing time $t_{E}$ by the source in the LMC ($%
D_{L}=25 $ kpc and $D_{S}=50$ kpc assumed). When the wormhole lens is bound
to the Galaxy, the transverse velocity $v_{T}$ of the lens relative to the
source star and observer is assumed to be $v_{T}=220$ km/s and when the lens
is unbound, the velocity is assumed to be $v_{T}=5000$ km/s. The charge $Q$
is scaled by $r_{\text{th}}$ under the restriction that $\frac{Q}{r_{\text{th}}}<\frac{1}{\sqrt{2}}=0.707$ so that $R_{E}$ is real [see Eq.(47)]}
\end{table}

Tables 1 \& 2 show the Einstein radii and angular Einstein radii for Bulge and LMC stars, for which we assumed the transverse velocity of stars $v_{T} = 220$ km/s and $v_{T} = 5000$ km/s for bound and unbound lenses respectively \cite{Abe:2010}. The detection of the magnification of star brightness depends on the observational timescale. Since the microlensing observations occur once every few hours, an event for which the timescale is small, say $t_{E}$ $<1$ day, is difficult to detect. Assuming a longer observational timescale $t_{E}\geq 1000$ days, and a total length of observation time about $10$ years, Abe \cite{Abe:2010} has argued for the Ellis-Bronnikov wormhole ($Q=0$) that the corresponding $r_{\text{th}}$ that one may detect should lie, for bound wormholes in the range $10^{2}$ km $\leq r_{\text{th}}\leq 10^{7}$ km and for unbound wormholes in the range $10^{5}$ km $\leq r_{\text{th}}\leq 10^{9}$ km obtained by using the $R_{E}$ and $v_{T}$ in Eqs.(47) and (86) respectively. In our case, $R_{E}$ depends not only on $r_{\text{th}}$ but also on charge $Q$ constrained by the limit $Q/r_{\text{th}}<1/\sqrt{2}$ [see Eq.(48)]. Realistically, assuming $Q$ to be too small such that $Q/r_{\text{th}}<<1/\sqrt{2}$, the limits on $r_{\text{th}}$ are not expected to practically differ much from the large values given above. Note that the detection of a lens with $R_{E}$ smaller than the star radius ($\simeq 10^{6} $ km), say $R_{E}<1$ km implying a tiny $r_{\text{th}}$, is very difficult because most of the features of the gravitational lensing are smeared out by the finite-source effect.

\begin{figure}[!ht]
  \centerline{\includegraphics[scale=0.35]{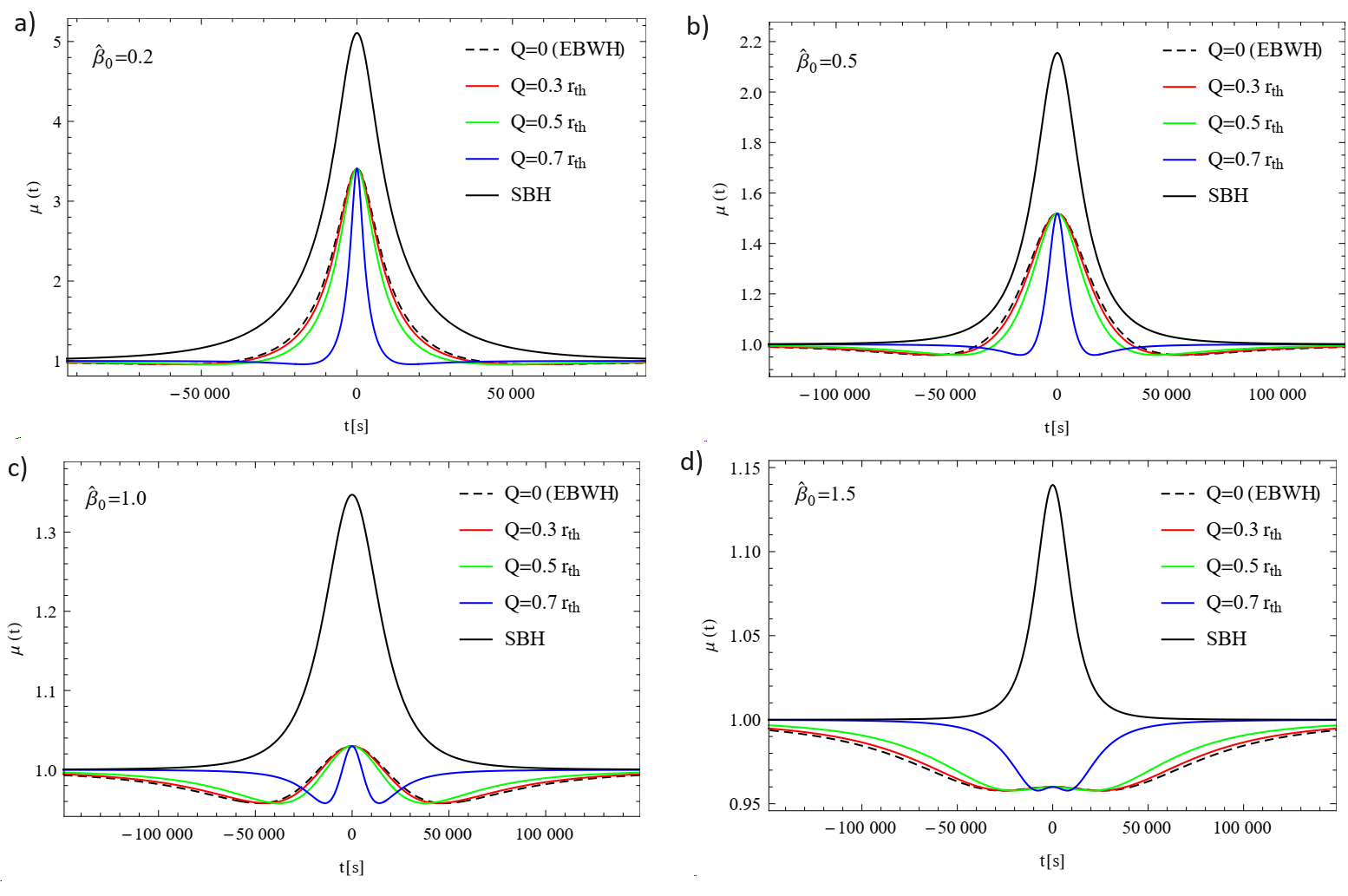}}
  \caption{The Paczy\'{n}ski light curves of Kim-Lee charged wormhole for $\hat{\beta}_{0}=0.2$ (a -- top left), $\hat{\beta}_{0}=0.5$ (b -- top right), $\hat{\beta}_{0}=1$ (c -- bottom left), and $\hat{\beta}_{0}=1.5$ (d -- bottom right). Various values of the $Q$ parameter, denoted as follows: $Q=0$ (massless Ellis-Bronnikov wormhole indicated by EBWH) -- dashed black line, $Q=0.3r_{\scriptsize{\textmd{th}}}$ -- solid red line, $Q=0.5r_{\scriptsize{\textmd{th}}}$ -- solid green line, $Q=0.7r_{\scriptsize{\textmd{th}}}$ -- solid blue line and solid black line corresponds Schwarzchild black hole (SBH).}
\end{figure}

Figs.7a-d show the light curves of charged wormhole for different values of the parameter $\hat{\beta}_{0}=0.2$ (a -- upper left), $0.5$ (b -- upper right), $1$ (c -- lower left) and $1.5$ (e -- lower right).

In detail, Fig.7a shows a plot of the charged wormhole light curve with a electric charge at $\hat{\beta}_{0} = 0.2$ for various values of the $Q$ parameter, denoted as follows: $Q=0$ (dashed black line), $0.3r_{\text{th}}$ (red line), $0.5r_{\text{th}}$ (green line) and $0.7r_{\text{th}}$ (blue line). As in the case of the Ellis-Bronnikov wormhole, the light curve contains two identical troughs on either side of the peak at any value of $Q$. The value of the valley is reduced by $4.2$ percent relative to the standard value, which is the same for all cases considered and is defined to be independent of $\hat{\beta}_{0}$. The peak value is $3.4$ times bigger than the standard value. Interestingly, the maximum and minimum brightness values of the star do not change when changing the parameter $Q$. However, from the figure we can see that as the parameter $Q$ increases, the troughs get closer to the peak, and the peak time itself decreases. This means that the larger the electric charge, the shorter the time for increasing brightness. Compared to a black hole, the light curve of a wormhole is much dimmer. Wormhole with electric charge at $\hat{\beta}_{0}=0.5$ is shown in Fig.7b. In this case, the peak and troughs of the light curve are also constant as the charge changes, and the magnitude of the peak is larger than the standard value by a factor of $1.52$. Just as in the previous case, an increase in the electric charge leads to the approach of the troughs towards each other meeting at the center. The same behavior is also observed in Fig.7c, the peak value is only $1.03$ times the standard value.

A more interesting case appears when $\hat{\beta}_{0}\geq 1$, for example we took the value $\hat{\beta}_{0}=1.5$ in Fig.7d. In this case, the peak of the light curve is located below the standard value by $\sim 4$\%. Thus, when $\hat{\beta}_{0}>1$, there will be no increase in the brightness of the source. On the contrary, it will look even dimmer than the illumination by the black hole.

%%%%%%%%%%%%%%%%%%%%%%%%%%%%%%%%%%%%%%%%%
\section{Probabilistic characteristics of microlensing}\label{probability}
%%%%%%%%%%%%%%%%%%%%%%%%%%%%%%%%%%%%%%%%%
The probability of a microlensing event to occur for a star is expressed by the optical depth $\tau $ \cite{Abe:2010}: 
\begin{equation}
\tau =\pi \int\limits_{0}^{D_{S}}n(D_{L})R_{E}^{2}dD_{L},
\end{equation}%
where $n(D_{L})$ is the number density of wormholes as a function of the line of sight. Here we simply assume that $n(D_{L})$ is constant ($n(D_{L})=n $): 
\begin{equation}
\tau =\pi n\int\limits_{0}^{D_{S}}\left[ \frac{\pi }{4}\frac{D_{L}\left(
D_{S}-D_{L}\right) \left( b_{0}^{2}-3Q^{2}\right) }{D_{S}}\right] ^{\frac{2}{%
3}}dD_{L}
\end{equation}%
and ussing new variable $x=D_{L}/D_{S}$ we can rewrite the above as 
\begin{equation}
\tau =2^{-\frac{4}{3}}\pi ^{\frac{5}{3}}nD_{S}^{\frac{5}{3}}\left(b_{0}^{2}-3Q^{2}\right) ^{\frac{2}{3}}\int\limits_{0}^{1}\left[ x\left(1-x\right) \right] ^{\frac{2}{3}}dx.
\end{equation}%
After integrating we get 
\begin{eqnarray}
\tau &=&\frac{3^{\frac{1}{2}}\pi ^{\frac{13}{6}}nD_{S}^{\frac{5}{3}}\left(
b_{0}^{2}-3Q^{2}\right) ^{\frac{2}{3}}\Gamma \left( \frac{5}{6}\right) }{%
7\cdot 4^{\frac{1}{3}}\Gamma \left( \frac{1}{3}\right) } \\
&\simeq &0.785nD_{S}^{\frac{5}{3}}\left( b_{0}^{2}-3Q^{2}\right) ^{\frac{2}{3%
}} \\
&=&\left( 1-\frac{2Q^{2}}{r_{\text{th}}^{2}}\right) r_{\text{th}}^{2},
\end{eqnarray}%
where $\Gamma \left( z\right) $ is the Euler gamma function.

The event rate expected for a source star $\widetilde{\Gamma}$ is defined by \cite{Abe:2010} 
\begin{equation}
\widetilde{\Gamma }=2\int\limits_{0}^{D_{S}}n(D_{L})R_{E}v_{T}dD_{L}.
\end{equation}%
Using the Einstein radius, given by Eq.(47), and assuming that $n(D_{L})=n$ is constant we can rewrite the above as 
\begin{equation}
\widetilde{\Gamma }=2\int\limits_{0}^{D_{S}}n\left[ \frac{\pi }{4}\frac{D_{L}\left( D_{S}-D_{L}\right) \left( b_{0}^{2}-3Q^{2}\right) }{D_{S}}\right]^{\frac{1}{3}}v_{T}dD_{L}
\end{equation}%
and ussing new variable $x=D_{L}/D_{S}$ we can rewrite the above as 
\begin{equation}
\widetilde{\Gamma }=2^{\frac{1}{3}}\pi ^{\frac{1}{3}}nv_{T}D_{S}^{\frac{4}{3}%
}\left( b_{0}^{2}-3Q^{2}\right) ^{\frac{1}{3}}\int\limits_{0}^{1}\left[
x\left( 1-x\right) \right] ^{\frac{1}{3}}dx.
\end{equation}%
After integrating we get 
\begin{eqnarray}
\widetilde{\Gamma } &=&\frac{2^{\frac{5}{3}}3^{\frac{1}{2}}\pi ^{\frac{11}{6}}}{5\Gamma \left( -\frac{1}{6}\right) \Gamma \left( \frac{2}{3}\right) } nv_{T}D_{S}^{\frac{4}{3}}\left( b_{0}^{2}-3Q^{2}\right) ^{\frac{1}{3}} \\
&\simeq &0.978nv_{T}D_{S}^{\frac{4}{3}}\left( b_{0}^{2}-3Q^{2}\right) ^{\frac{1}{3}} \\
&=&0.978nv_{T}D_{S}^{\frac{4}{3}}\left( 1-\frac{2Q^{2}}{r_{\text{th}}^{2}}\right) r_{\text{th}}^{2}.
\end{eqnarray}

To show the numerical values of optical depth and event rate for the charged wormhole, we will the use numerology presented by Abe \cite{Abe:2010}. For wormholes bound to the Galaxy, the number density is approximately equal to $n=\rho _{\text{Ls}}/M_{\text{star}}=0.147$ pc$^{-3}$, where $\rho _{\text{Ls}}$ is the local stellar density in the solar neighborhood, $\rho _{\text{Ls}} = 0.044M_{\text{Sun}}$ pc$^{-3}$, and $M_{\text{star}}$ is the average mass of stars taken as $M_{\text{star}} = 0.3M_{\text{Sun}}$, a typical mass of a dominant stellar component in the Galaxy such as M-dwarf. Using these values, we calculated the optical depths and event rates for galactic Bulge lensing presented in Table 5 and for LMC lensing, in Table 6.

\begin{table}
\begin{tabular}{|c|c|c|c||c|c|}
\hline
&  & \multicolumn{2}{|c||}{Bound} & \multicolumn{2}{|c|}{Unbound} \\ 
%\cline{3-6}
$r_{\text{th}}$ [km] & $Q/r_{\text{th}}$ & $\tau $ & $\widetilde{\Gamma }$
[1/year] & $\tau $ & $\widetilde{\Gamma }$ [1/year] \\ \hline
$10^{0}$ & $0$ & $3.8165\times 10^{-13}$ & $5.2735\times 10^{-9}$ & $%
1.2903\times 10^{-20}$ & $4.0522\times 10^{-15}$ \\ 
$10^{0}$ & $0.3$ & $3.3435\times 10^{-13}$ & $4.6200\times 10^{-9}$ & $%
1.1304\times 10^{-20}$ & $3.5500\times 10^{-15}$ \\ 
$10^{0}$ & $0.5$ & $2.4042\times 10^{-13}$ & $3.3221\times 10^{-9}$ & $%
8.1286\times 10^{-21}$ & $2.5527\times 10^{-15}$ \\ 
$10^{0}$ & $0.7$ & $2.8120\times 10^{-14}$ & $3.8856\times 10^{-10}$ & $%
9.5073\times 10^{-22}$ & $2.9857\times 10^{-16}$ \\ \hline
$10^{1}$ & $0$ & $8.2224\times 10^{-12}$ & $2.4478\times 10^{-8}$ & $%
2.7799\times 10^{-19}$ & $1.8808\times 10^{-14}$ \\ 
$10^{1}$ & $0.3$ & $7.2034\times 10^{-12}$ & $2.1444\times 10^{-8}$ & $%
2.4354\times 10^{-19}$ & $1.6478\times 10^{-14}$ \\ 
$10^{1}$ & $0.5$ & $5.1798\times 10^{-12}$ & $1.5420\times 10^{-8}$ & $%
1.7512\times 10^{-19}$ & $1.1849\times 10^{-14}$ \\ 
$10^{1}$ & $0.7$ & $6.0583\times 10^{-13}$ & $1.8035\times 10^{-9}$ & $%
2.0483\times 10^{-20}$ & $1.3858\times 10^{-15}$ \\ \hline
$10^{2}$ & $0$ & $1.7714\times 10^{-10}$ & $1.1361\times 10^{-7}$ & $%
5.9892\times 10^{-18}$ & $8.7301\times 10^{-14}$ \\ 
$10^{2}$ & $0.3$ & $1.5519\times 10^{-10}$ & $9.9535\times 10^{-8}$ & $%
5.2470\times 10^{-18}$ & $7.6483\times 10^{-14}$ \\ 
$10^{2}$ & $0.5$ & $1.1159\times 10^{-10}$ & $7.1573\times 10^{-8}$ & $%
3.7730\times 10^{-18}$ & $5.4996\times 10^{-14}$ \\ 
$10^{2}$ & $0.7$ & $1.3052\times 10^{-11}$ & $8.3712\times 10^{-9}$ & $%
4.4129\times 10^{-19}$ & $6.4324\times 10^{-15}$ \\ \hline
$10^{3}$ & $0$ & $3.8165\times 10^{-9}$ & $5.2735\times 10^{-7}$ & $%
1.2903\times 10^{-16}$ & $4.0522\times 10^{-13}$ \\ 
$10^{3}$ & $0.3$ & $3.3435\times 10^{-9}$ & $4.6200\times 10^{-7}$ & $%
1.1304\times 10^{-16}$ & $3.5500\times 10^{-13}$ \\ 
$10^{3}$ & $0.5$ & $2.4042\times 10^{-9}$ & $3.3221\times 10^{-7}$ & $%
8.1286\times 10^{-17}$ & $2.5527\times 10^{-13}$ \\ 
$10^{3}$ & $0.7$ & $2.8120\times 10^{-10}$ & $3.8856\times 10^{-8}$ & $%
9.5073\times 10^{-18}$ & $2.9857\times 10^{-14}$ \\ \hline
$10^{4}$ & $0$ & $8.2224\times 10^{-8}$ & $2.4478\times 10^{-6}$ & $%
2.7799\times 10^{-15}$ & $1.8808\times 10^{-12}$ \\ 
$10^{4}$ & $0.3$ & $7.2034\times 10^{-8}$ & $2.1444\times 10^{-6}$ & $%
2.4354\times 10^{-15}$ & $1.6478\times 10^{-12}$ \\ 
$10^{4}$ & $0.5$ & $5.1798\times 10^{-8}$ & $1.5420\times 10^{-6}$ & $%
1.7512\times 10^{-15}$ & $1.1849\times 10^{-12}$ \\ 
$10^{1}$ & $0.7$ & $6.0583\times 10^{-9}$ & $1.8035\times 10^{-7}$ & $%
2.0483\times 10^{-16}$ & $1.3858\times 10^{-13}$ \\ \hline
$10^{5}$ & $0$ & $1.7714\times 10^{-5}$ & $1.1361\times 10^{-5}$ & $%
5.9892\times 10^{-14}$ & $8.7301\times 10^{-12}$ \\ 
$10^{5}$ & $0.3$ & $1.5519\times 10^{-6}$ & $9.9535\times 10^{-6}$ & $%
5.2470\times 10^{-14}$ & $7.6483\times 10^{-12}$ \\ 
$10^{5}$ & $0.5$ & $1.1159\times 10^{-6}$ & $7.1573\times 10^{-6}$ & $%
3.7730\times 10^{-14}$ & $5.4996\times 10^{-12}$ \\ 
$10^{5}$ & $0.7$ & $1.3052\times 10^{-7}$ & $8.3712\times 10^{-7}$ & $%
4.4129\times 10^{-15}$ & $6.4324\times 10^{-13}$ \\ \hline
$10^{6}$ & $0$ & $3.8165\times 10^{-5}$ & $5.2735\times 10^{-5}$ & $%
1.2903\times 10^{-12}$ & $4.0522\times 10^{-11}$ \\ 
$10^{6}$ & $0.3$ & $3.3435\times 10^{-5}$ & $4.6200\times 10^{-5}$ & $%
1.1304\times 10^{-12}$ & $3.5500\times 10^{-11}$ \\ 
$10^{6}$ & $0.5$ & $2.4042\times 10^{-5}$ & $3.3221\times 10^{-5}$ & $%
8.1286\times 10^{-13}$ & $2.5527\times 10^{-11}$ \\ 
$10^{6}$ & $0.7$ & $2.8120\times 10^{-6}$ & $3.8856\times 10^{-6}$ & $%
9.5073\times 10^{-14}$ & $2.9857\times 10^{-12}$ \\ \hline
$10^{7}$ & $0$ & $8.2224\times 10^{-4}$ & $2.4478\times 10^{-4}$ & $%
2.7799\times 10^{-11}$ & $1.8808\times 10^{-10}$ \\ 
$10^{7}$ & $0.3$ & $7.2034\times 10^{-4}$ & $2.1444\times 10^{-4}$ & $%
2.4354\times 10^{-11}$ & $1.6478\times 10^{-10}$ \\ 
$10^{7}$ & $0.5$ & $5.1798\times 10^{-4}$ & $1.5420\times 10^{-4}$ & $%
1.7512\times 10^{-11}$ & $1.1849\times 10^{-10}$ \\ 
$10^{7}$ & $0.7$ & $6.0583\times 10^{-5}$ & $1.8035\times 10^{-5}$ & $%
2.0483\times 10^{-12}$ & $1.3858\times 10^{-11}$ \\ \hline
$10^{8}$ & $0$ & $1.7714\times 10^{-2}$ & $1.1361\times 10^{-3}$ & $%
5.9892\times 10^{-10}$ & $8.7301\times 10^{-10}$ \\ 
$10^{8}$ & $0.3$ & $1.5519\times 10^{-2}$ & $9.9535\times 10^{-4}$ & $%
5.2470\times 10^{-10}$ & $7.6483\times 10^{-10}$ \\ 
$10^{8}$ & $0.5$ & $1.1159\times 10^{-2}$ & $7.1573\times 10^{-4}$ & $%
3.7730\times 10^{-10}$ & $5.4996\times 10^{-10}$ \\ 
$10^{8}$ & $0.7$ & $1.3052\times 10^{-3}$ & $8.3712\times 10^{-5}$ & $%
4.4129\times 10^{-11}$ & $6.4324\times 10^{-11}$ \\ \hline
$10^{9}$ & $0$ & $3.8165\times 10^{-1}$ & $5.2735\times 10^{-3}$ & $%
1.2903\times 10^{-8}$ & $4.0522\times 10^{-9}$ \\ 
$10^{9}$ & $0.3$ & $3.3435\times 10^{-1}$ & $4.6200\times 10^{-3}$ & $%
1.1304\times 10^{-8}$ & $3.5500\times 10^{-9}$ \\ 
$10^{9}$ & $0.5$ & $2.4042\times 10^{-1}$ & $3.3221\times 10^{-3}$ & $%
8.1286\times 10^{-9}$ & $2.5527\times 10^{-9}$ \\ 
$10^{9}$ & $0.7$ & $2.8120\times 10^{-2}$ & $3.8856\times 10^{-4}$ & $%
9.5073\times 10^{-10}$ & $2.9857\times 10^{-10}$ \\ \hline
$10^{10}$ & $0$ & $8.2224$ & $2.4478\times 10^{-2}$ & $2.7799\times 10^{-7}$
& $1.8808\times 10^{-8}$ \\ 
$10^{10}$ & $0.3$ & $7.2034$ & $2.1444\times 10^{-2}$ & $2.4354\times
10^{-7} $ & $1.6478\times 10^{-8}$ \\ 
$10^{10}$ & $0.5$ & $5.1798$ & $1.5419\times 10^{-2}$ & $1.7512\times
10^{-7} $ & $1.1849\times 10^{-8}$ \\ 
$10^{10}$ & $0.7$ & $0.6058$ & $1.8035\times 10^{-3}$ & $2.0483\times
10^{-8} $ & $1.3858\times 10^{-9}$ \\ \hline
\end{tabular}
\caption{The optical depth $\tau $ and event rate $\widetilde{\Gamma}$ of a
microlensing event by the charged wormhole, when the source star is in the
Bulge ($D_{L}=4$ kpc and $D_{S}=8$ kpc assumed). When the wormhole lens is
bound to the Galaxy, the velocity of source $v_{T}=220$ km/s and number
density of wormholes $n=0.147$ pc$^{-3}$ are assumed \cite{Abe:2010} and when unbound, $%
v_{T}=5000$ km/s and $n=4.97\times 10^{-9}$ pc$^{-3}$ are assumed \cite{Abe:2010}. The
charge\ $Q$ is scaled by $r_{\text{th}}$ under the restriction that $\frac{Q%
}{r_{\text{th}}}<\frac{1}{\sqrt{2}}=0.707$ so that $R_{E}$ is real [see
Eq.(47)].}
\end{table}

\begin{table}
\begin{tabular}{|c|c|c|c||c|c|}
\hline
&  & \multicolumn{2}{|c||}{Bound} & \multicolumn{2}{|c|}{Unbound} \\ 
%\cline{3-6}
$r_{\text{th}}$ [km] & $Q/r_{\text{th}}$ & $\tau $ & $\widetilde{\Gamma }$
[1/year] & $\tau $ & $\widetilde{\Gamma }$ [1/year] \\ \hline
$10^{0}$ & $0$ & $8.0934\times 10^{-12}$ & $6.0712\times 10^{-8}$ & $%
2.7363\times 10^{-19}$ & $4.6651\times 10^{-14}$ \\ 
$10^{0}$ & $0.3$ & $7.0904\times 10^{-12}$ & $5.3188\times 10^{-8}$ & $%
2.3972\times 10^{-19}$ & $4.0870\times 10^{-14}$ \\ 
$10^{0}$ & $0.5$ & $5.0985\times 10^{-12}$ & $3.8246\times 10^{-8}$ & $%
1.7238\times 10^{-19}$ & $2.9388\times 10^{-14}$ \\ 
$10^{0}$ & $0.7$ & $5.9633\times 10^{-13}$ & $4.4733\times 10^{-9}$ & $%
2.0161\times 10^{-20}$ & $3.4373\times 10^{-15}$ \\ \hline
$10^{1}$ & $0$ & $1.7437\times 10^{-10}$ & $2.8180\times 10^{-7}$ & $%
5.8952\times 10^{-18}$ & $2.1653\times 10^{-13}$ \\ 
$10^{1}$ & $0.3$ & $1.5276\times 10^{-10}$ & $2.4688\times 10^{-7}$ & $%
5.1647\times 10^{-18}$ & $1.8970\times 10^{-13}$ \\ 
$10^{1}$ & $0.5$ & $1.0984\times 10^{-10}$ & $1.7752\times 10^{-7}$ & $%
3.7138\times 10^{-18}$ & $1.3641\times 10^{-13}$ \\ 
$10^{1}$ & $0.7$ & $1.2847\times 10^{-11}$ & $2.0763\times 10^{-8}$ & $%
4.3437\times 10^{-19}$ & $1.5954\times 10^{-14}$ \\ \hline
$10^{2}$ & $0$ & $3.7566\times 10^{-9}$ & $1.3080\times 10^{-6}$ & $%
1.2701\times 10^{-16}$ & $1.0051\times 10^{-12}$ \\ 
$10^{2}$ & $0.3$ & $3.2911\times 10^{-9}$ & $1.1459\times 10^{-6}$ & $%
1.1127\times 10^{-16}$ & $8.8052\times 10^{-13}$ \\ 
$10^{2}$ & $0.5$ & $2.3665\times 10^{-9}$ & $8.2399\times 10^{-7}$ & $%
8.0011\times 10^{-17}$ & $6.3315\times 10^{-13}$ \\ 
$10^{2}$ & $0.7$ & $2.7679\times 10^{-10}$ & $9.6374\times 10^{-8}$ & $%
9.3581\times 10^{-18}$ & $7.4054\times 10^{-14}$ \\ \hline
$10^{3}$ & $0$ & $8.0934\times 10^{-8}$ & $6.0712\times 10^{-6}$ & $%
2.7363\times 10^{-15}$ & $4.6651\times 10^{-12}$ \\ 
$10^{3}$ & $0.3$ & $7.0904\times 10^{-8}$ & $5.3188\times 10^{-6}$ & $%
2.3972\times 10^{-15}$ & $4.0870\times 10^{-12}$ \\ 
$10^{3}$ & $0.5$ & $5.0985\times 10^{-8}$ & $3.8246\times 10^{-6}$ & $%
1.7238\times 10^{-15}$ & $2.9388\times 10^{-12}$ \\ 
$10^{3}$ & $0.7$ & $5.9633\times 10^{-9}$ & $4.4733\times 10^{-7}$ & $%
2.0161\times 10^{-16}$ & $3.4373\times 10^{-13}$ \\ \hline
$10^{4}$ & $0$ & $1.7437\times 10^{-6}$ & $2.8180\times 10^{-5}$ & $%
5.8952\times 10^{-14}$ & $2.1653\times 10^{-11}$ \\ 
$10^{4}$ & $0.3$ & $1.5276\times 10^{-6}$ & $2.4688\times 10^{-5}$ & $%
5.1647\times 10^{-14}$ & $1.8970\times 10^{-11}$ \\ 
$10^{4}$ & $0.5$ & $1.0984\times 10^{-6}$ & $1.7752\times 10^{-5}$ & $%
3.7138\times 10^{-14}$ & $1.3641\times 10^{-11}$ \\ 
$10^{4}$ & $0.7$ & $1.2847\times 10^{-7}$ & $2.0763\times 10^{-6}$ & $%
4.3437\times 10^{-15}$ & $1.5954\times 10^{-12}$ \\ \hline
$10^{5}$ & $0$ & $3.7566\times 10^{-5}$ & $1.3080\times 10^{-4}$ & $%
1.2701\times 10^{-12}$ & $1.0051\times 10^{-10}$ \\ 
$10^{5}$ & $0.3$ & $3.2911\times 10^{-5}$ & $1.1459\times 10^{-4}$ & $%
1.1127\times 10^{-12}$ & $8.8052\times 10^{-11}$ \\ 
$10^{5}$ & $0.5$ & $2.3665\times 10^{-5}$ & $8.2399\times 10^{-5}$ & $%
8.0011\times 10^{-13}$ & $6.3315\times 10^{-11}$ \\ 
$10^{5}$ & $0.7$ & $2.7679\times 10^{-6}$ & $9.6374\times 10^{-6}$ & $%
9.3581\times 10^{-14}$ & $7.4054\times 10^{-12}$ \\ \hline
$10^{6}$ & $0$ & $8.0934\times 10^{-4}$ & $6.0712\times 10^{-4}$ & $%
2.7363\times 10^{-11}$ & $4.6651\times 10^{-10}$ \\ 
$10^{6}$ & $0.3$ & $7.0904\times 10^{-4}$ & $5.3188\times 10^{-4}$ & $%
2.3972\times 10^{-11}$ & $4.0870\times 10^{-10}$ \\ 
$10^{6}$ & $0.5$ & $5.0985\times 10^{-4}$ & $3.8246\times 10^{-4}$ & $%
1.7238\times 10^{-11}$ & $2.9388\times 10^{-10}$ \\ 
$10^{6}$ & $0.7$ & $5.9633\times 10^{-5}$ & $4.4733\times 10^{-5}$ & $%
2.0161\times 10^{-12}$ & $3.4373\times 10^{-11}$ \\ \hline
$10^{7}$ & $0$ & $1.7437\times 10^{-2}$ & $2.8180\times 10^{-3}$ & $%
5.8952\times 10^{-10}$ & $2.1653\times 10^{-9}$ \\ 
$10^{7}$ & $0.3$ & $1.5276\times 10^{-2}$ & $2.4688\times 10^{-3}$ & $%
5.1647\times 10^{-10}$ & $1.8970\times 10^{-9}$ \\ 
$10^{7}$ & $0.5$ & $1.0984\times 10^{-2}$ & $1.7752\times 10^{-3}$ & $%
3.7138\times 10^{-10}$ & $1.3641\times 10^{-9}$ \\ 
$10^{7}$ & $0.7$ & $1.2847\times 10^{-3}$ & $2.0763\times 10^{-4}$ & $%
4.3437\times 10^{-11}$ & $1.5954\times 10^{-10}$ \\ \hline
$10^{8}$ & $0$ & $3.7566\times 10^{-1}$ & $1.3080\times 10^{-2}$ & $%
1.2701\times 10^{-8}$ & $1.0051\times 10^{-8}$ \\ 
$10^{8}$ & $0.3$ & $3.2911\times 10^{-1}$ & $1.1459\times 10^{-2}$ & $%
1.1127\times 10^{-8}$ & $8.8052\times 10^{-9}$ \\ 
$10^{8}$ & $0.5$ & $2.3665\times 10^{-1}$ & $8.2399\times 10^{-3}$ & $%
8.0011\times 10^{-9}$ & $6.3315\times 10^{-9}$ \\ 
$10^{8}$ & $0.7$ & $2.7679\times 10^{-2}$ & $9.6374\times 10^{-4}$ & $%
9.3581\times 10^{-10}$ & $7.4054\times 10^{-10}$ \\ \hline
$10^{9}$ & $0$ & $8.0934$ & $6.0712\times 10^{-2}$ & $2.7363\times 10^{-7}$
& $4.6651\times 10^{-8}$ \\ 
$10^{9}$ & $0.3$ & $7.0904$ & $5.3188\times 10^{-2}$ & $2.3972\times 10^{-7}$
& $4.0870\times 10^{-8}$ \\ 
$10^{9}$ & $0.5$ & $5.0985$ & $3.8246\times 10^{-2}$ & $1.7238\times 10^{-7}$
& $2.9388\times 10^{-8}$ \\ 
$10^{9}$ & $0.7$ & $0.5963$ & $4.4733\times 10^{-3}$ & $2.0161\times 10^{-8}$
& $3.4373\times 10^{-9}$ \\ \hline
$10^{7}$ & $0$ & $174.3667$ & $2.8180\times 10^{-1}$ & $5.8952\times 10^{-6}$
& $2.1653\times 10^{-7}$ \\ 
$10^{7}$ & $0.3$ & $152.7587$ & $2.4688\times 10^{-1}$ & $5.1647\times
10^{-6}$ & $1.8970\times 10^{-7}$ \\ 
$10^{7}$ & $0.5$ & $109.8441$ & $1.7752\times 10^{-1}$ & $3.7138\times
10^{-6}$ & $1.3641\times 10^{-7}$ \\ 
$10^{7}$ & $0.7$ & $12.8474$ & $2.0763\times 10^{-2}$ & $4.3437\times
10^{-7} $ & $1.5954\times 10^{-8}$ \\ \hline
\end{tabular}
\caption{The optical depth $\tau $ and event rate $\widetilde{\Gamma }$ of a
microlensing event by the charged wormhole, when the source star is in the
LMC ($D_{L}=25$ kpc and $D_{S}=50$ kpc assumed). When the wormhole lens is
bound to the Galaxy, the transverse velocity of source $v_{T}=220$ km/s and
number density of wormholes $n=0.147$ pc$^{-3}$ are assumed \cite{Abe:2010} and when
unbound, the velocity of source $v_{T}=5000$ km/s and number density of
wormholes $n=4.97\times 10^{-9}$ pc$^{-3}$ are assumed \cite{Abe:2010}. The charge $Q$
is scaled by $r_{\text{th}}$ under the restriction that $\frac{Q}{r_{\text{th}}}<\frac{1}{\sqrt{2}}=0.707$ so that $R_{E}$ is real [see Eq.(47)].}
\end{table}

Some speculative remarks are in order. In an ordinary Schwarzschild microlensing survey, observations are made of more than $10$ million stars. Thus, we can expect approximately $10^{7}\widetilde{\Gamma}$ events in a year. However, the situation is different in a wormhole search by studying their lensing signatures. One of the complicating reason is that the magnification of wormhole lensing falls below that of Schwarzschild lensing, and that there is a decrease in brightness (gutters) around the Einstein radius crossing times. On the other hand, microlensing surveys so far have mainly searched for source stars that increase in brightness. We need to find lensed stars that decrease in brightness in the wormhole search, which may be compensated by the brighter stars. This means that, in the wormhole search, far fewer stars can be monitored than in an ordinary microlensing survey. Furthermore, low magnification in wormhole lensing reduces the detection efficiency of the wormhole. Considering these complications, it is assumed \cite{Abe:2010} that the effective number of stars that need to be monitored to find a wormhole is $10^{6}$. Therefore, to expect more than one event in a survey of several years, $\widetilde{\Gamma }$ must be greater than $10^{-6}$. The values in Table 5 indicate that the detection of wormholes with $r_{\text{th}}\geq 10^{4}$ km is expected in the microlensing survey of the Galactic bulge stars in the case of the bound lens model. Similar arguments apply to LMC lensing as presented in Table 6. and we expect $\widetilde{\Gamma}>10^{-6}$ to find a wormhole.

%%%%%%%%%%%%%%%%%%%%%%%%%%%%%%%%%%%%%%%%%
\section{Conclusion}\label{concl}
%%%%%%%%%%%%%%%%%%%%%%%%%%%%%%%%%%%%%%%%%
The lensing phenomenon by wormholes is an active area of current research that can help detecting the nature of the lens from the observable lensing signatures (see, e.g., \cite{Tsukamoto:2012}). The present work considered new types of charged wormholes, proposed by Kim and Lee \cite{Kim:2001}, resulting from the effect of backreaction of additional matter on the background metric (5) - one type is scalar charged (9) and the other is electrically charged wormhole (20). These wormholes could provide novel applications to galactic microlensing, when the configuration is not clean enough, that is, when additional matter-energy distribution surrounds the lens, for instance, dust or some energy field. This matter has been taken into account in the backreaction problem leading to Kim-Lee wormholes as exact solutions. In their role as lenses, they would then exhibit the effect of additional matter via lensing signatures of the backreacted metric geometry. To our knowledge, the idea of using backreacted metric as a lens for estimating the influence of additional matter on lensing scenario is the first of its kind that unequivocally provides information about the nature of the backreacted lens.

In the present work, we considered the equation of state parameter $\gamma = -1$, studied by Kim and Lee, representing the background zero mass Ellis-Bronnikov wormhole. It is found that the ADM masses of both the background and backreacted spacetimes are zero even though their redshift and/or shape functions differ. We observed that the additional scalar charge $\alpha$ only shifts the background shape function $b_{\text{shape}}(r) = b_{0}^{2}/r$ to $B_{\text{shape}}(r) = \left( b_{0}^{2}-\alpha \right) /r\equiv a^{2}/r$ but the corresponding source scalar fields, $\varphi _{\text{KL}}$ and $\varphi _{\text{EB}}$, differ in general. This shift is a result of linear superposition via field equations of same types of matter distribution, viz., the minimally coupled scalar field. Microlensing for the background metric ($\Phi =0$) without additional matter has already been studied by Abe \cite{Abe:2010}. So here we focused on the electrically charged wormhole which, even though massless, admits an extra degree of freedom, the electric charge $Q$, that appears in the redshift function ($\Phi \neq 0$) and studied the effect of $Q\neq 0$ on Galactic microlensing characteristics. Our objective was to compare the microlensing profiles of electrically charged wormhole with the profiles of the background massless Ellis-Bronnikov wormhole and a Schwarzschild black hole to conclude about the lens in question.

We assumed the charged wormholes to be halo objects lensing the source stars belonging to the Galactic Bulge and to the Large Magellanic Cloud (LMC). The distances involved between the observer (on Earth), the lens and the source stars suggest that the concept of microlensing is well justified since the Einstein angle is too tiny, $\theta _{E}\approx 10^{-5\prime \prime }$ to $10^{-4\prime \prime }$. This fact allowed us to use the PPN expansion method of Keeton and Petters \cite{Keeton:2005} to find the effect of $Q$ on the lensing observables such as the image position, magnification, centroid, light curves and time delay of images for source stars. Additionally, the probabilistic features such as optical depth and event rate are calculated based on the hypothesis that the wormhole lens could be bound or unbound to the Galaxy.

Qualitative generic effects of $Q\neq 0$ on lensing observables are as follows:

1. The presence of $Q$ reduces the deflection angle of light compared to the background Ellis-Bronnikov metric (EB), i.e., $\widehat{\alpha }_{\text{El}}(b) < \widehat{\alpha}_{\text{EB}}(b)$.

2. The positivity of throat radius $r_{\text{th}}$ constrains $Q < b_{0}$. However, the reality of Einstein radius $R_{E}$ provides a stronger constraint $Q < b_{0}/\sqrt{3}$, which translates to $\frac{Q}{r_{\text{th}}} < \frac{1}{\sqrt{2}}$. In the extreme case, $\frac{Q}{r_{\text{th}}}=\frac{1}{\sqrt{2}}$, all corrections appear to be zero, so the images should be sharp.

3. The zeroth order image position is independent of $Q$ and increases indefinitely regardless of the parity of images as is evident from Fig.3a. The correction to image positions lensed by the Ellis-Bronnikov wormhole is always greater in magnitude than the ones lensed by the electrically charged wormhole, as evident from Eqs.(55,56). It can be seen from the Fig.3b that within the canonical constraint $\frac{Q}{r_{\text{th}}}<\frac{1}{\sqrt{2}}$, the first order correction $\theta _{1}^{\text{(El)}}$ to the image position for the negative parity image ($\beta <0$) and for the positive parity image ($\beta >0$) are positive and symmetric around the lens-observer optical axis ($\beta =0$). The extreme case $Q=b_{0}/\sqrt{3}$ or $\frac{Q}{r_{\text{th}}} = \frac{1}{\sqrt{2}}$ gives $\theta _{1}^{\text{(El)}} = 0$, which means that there is no correction to the zeroth order image position. For $Q=0$, one has the zero mass Ellis-Bronnikov wormhole, for which the corrections are maximal. The plots for $\theta _{2}^{\text{(El)}}$ show that they are positive for $\beta <0$ and negative for $\beta >0$.

4. The magnification $\mu _{0}^{\text{(El)}} = 1$ and its correction $\mu_{2}^{\text{(El)}} = 0$ but $\mu _{1}^{\text{(El)}}$ behaves like $\theta_{1}^{\text{(El)}}$ (Figs.4). Also total magnification $\mu _{\text{tot}, 0}^{\text{(El)}} = 1,\mu _{\text{tot}, 1}^{\text{(El)}} = 0$ (Figs.5). The centroid $\Theta _{\text{cent}, 0}^{\text{(El)}} = |\beta |,$ $\Theta _{\text{cent}, 1}^{\text{(El)}} = 0$ and the second order correction to it $\Theta _{\text{cent}, 2}^{\text{(El)}} < 0$ for both parities (Figs.6).

5. The time difference in the arrival times from two distinct images is a useful diagnostic. However, practically observable quantity is the differential time delay between the positive and negative parity images $\hat{\tau}_{+}$ and $\hat{\tau}_{-}$ defined by $\Delta \hat{\tau}=\hat{\tau}_{-}-\hat{\tau}_{+}$. A remarkable result is that time delay $\Delta \hat{\tau}$ up to $O\left( {\varepsilon }\right) ^{2}$ is independent of the charge $Q$ of wormhole.

6. Table 1 shows the Einstein radii $R_{E}$ and angular Einstein radii $\theta _{E}$ for a Bulge star for which the assumed distances are $D_{S}=8$ kpc, $D_{L}=4$ kpc and the transverse velocity of stars are assumed to be $v_{T}=220$ km/s (bound wormhole) and $v_{T}=5000$ km/s (unbound wormhole). Table 2 shows Einstein radii $R_{E}$ and angular Einstein radii $\theta _{E}$, when the source is in the LMC ($D_{L}=25$ kpc and $D_{S}=50$ kpc assumed). From both the tables, it is seen that both $R_{E}$ and $\theta_{E} $ monotonically decrease as $Q/r_{\text{th}}$ increases towards the limiting value $0.7$. Table 3 shows the Einstein radius crossing time by the source, when it is in the Bulge ($D_{L}=4$ kpc and $D_{S}=8$ kpc) and the lens could be bound or unbound. Table 4 shows the Einstein radius crossing time by the source, when it is in the LMC and the lens could be bound or unbound. From both the tables, it is seen that the Einstein radius crossing time decreases with increase in $Q/r_{\text{th}}$ and the decrease becomes more rapid as $Q/r_{\text{th}}$ nears its limiting value $0.7$. These features are consistently reflected in the light curves.

7. The Paczy\'{n}ski light curves of Figs.7a-d show some interesting behavior that is characteristic of charged wormholes. Compared to the Schwarzschild black hole, the light curve of a wormhole is much dimmer. Also the light curves by the charged wormholes show gutters (brightness below base value $1$ occurring in the absence of lens) at the times the source enters and exits the Einstein ring. This feature alone, which is absent in Schwarzschild lensing, could enable us to detect the lens to be a massless wormhole by looking for its signatures in the past observed data. In detail, Figs.7a-d show the light curves for different values of the parameter $\hat{\beta}_{0}=0.2$ (a -- upper left), $0.5$ (b -- upper right), $1$ (c -- lower left) and $1.5$ (e -- lower right). Fig.7a shows light curve at $\hat{\beta}_{0}=0.2$ depends on various values of the charge $Q$ as follows: $Q=0$ (dashed black line), $0.3r_{\text{th}}$ (red line), $0.5r_{\text{th}}$ (green line), $0.7r_{\text{th}}$ (blue line) and light curves for Schwarzschild lensing are indicated by black curves. The light curves of charged wormholes show two identical gutters on either side of the peak at any value of parameter $Q$. The lowest depth (trough) of the gutter is less than $4.2$\% from the base value, which turns out to be the same for all cases considered, i.e. does not depend on $\hat{\beta}_{0}$. The peak value is $3.4$ times bigger than the base value. For a given $\hat{\beta}_{0}$, the maximum and minimum brightness values of the source star do not change when changing the parameter $Q$. \textit{However, from the figures, we can in general see that as the parameter $Q$ increases towards the extreme value $\frac{r_{\text{th}}}{\sqrt{2}}$, the two troughs of the gutters (lowest points on the blue lines) move closer from either side until the Einstein ring radius $R_{E}$ shrinks to zero allowing the source star to cross the point radius instantly at $t=0$. This feature alone is an intriguingly new phenomenon charaterizing an extreme charged Kim-Lee wormhole.} Since the crossing time is controlled by charge, the non-zero ring passage time may also potentially provide information on $Q$.

Paczy\'{n}ski light curves for charged wormhole at $\hat{\beta}_{0}=0.5$ are shown in Fig.7b. In this case, the peak and troughs of the light curves are constant as the charge varies, and the magnitude of the peak is larger than the base value by a factor of $1.52$. Just as in the previous case, an increase in the charge $Q$ leads to instant crossing. The same behavior is also observed in Fig.7c, only the peak value is only $1.03$ times the base value. A more interesting case appears when $\hat{\beta}_{0}\geq 1$, for example we took the value $\hat{\beta}_{0}=1.5$ in Fig.7d in which case, the gutters are located below $4$\% of the base value. Thus, when $\hat{\beta}_{0}>1$, there will be no increase in the brightness of the source, but on the contrary, it will look even dimmer than the base illumination. Any of the above microlensing features can provide signatures that can potentially reveal the existence of an electrically charged wormhole lens.

The probability of detecting a charged wormhole lens in the Bulge and in the LMC lensing has been briefly speculated in view of optical depth $\tau$ and event rates $\widetilde{\Gamma}$ in Sec.6. Their values, as tabulated in Table 5, indicate that the detection of wormholes with $r_{\text{th}}\geq 10^{4}$ km is expected in the microlensing survey of the Galactic bulge stars in the case of the bound lens model. Similar arguments apply to LMC lensing as presented in Table 6 and we expect $\widetilde{\Gamma}>10^{-6}$ to find at least one wormhole lens.

\section*{Acknowledgments}

This work was supported by the Russian Science Foundation under grant no. 23-22-00391, https://rscf.ru/en/project/23-22-00391/. We thank the two anonymous reviewers for their helpful comments.

\end{document}